# A generalized framework for the critical current for a one-dimensional crossed-field gap in an orthogonal coordinate system

Jack K. Wright[1,2], N.R. Sree Harsha[2,a)], and Allen L. Garner[2,3,4,b)]

[1]School of Aeronautics and Astronautics, Purdue University, West Lafayette, Indiana 47907, USA

[2]School of Nuclear Engineering, Purdue University, West Lafayette, Indiana 47907, USA

[3]Elmore Family School of Electrical and Computer Engineering, Purdue University, West Lafayette, Indiana 47907, USA

[4]Department of Agricultural and Biological Engineering, West Lafayette, Indiana 47907, USA

[a)]**Present address:** Department of Electrical and Computer Engineering, University of Rochester, Rochester, New York 14620, USA

[b)]**Author to whom correspondence should be addressed:** algarner@purdue.edu

Recent studies have applied variational calculus, conformal mapping, and point transformations to extend the one-dimensional SCLC from planar gaps to more complicated geometries. However, introducing a magnetic field orthogonal to the diode's electric field complicates these calculations due to changes in the electron trajectory. This paper extends a recent study that applied variational calculus to determine the SCLC for a cylindrical crossed-field diode to develop a framework that is valid for any orthogonal coordinate system. We then derive equations for the SCLC for crossed-field gaps in spherical, tip-to-tip, and tip-to-plane geometries that can be solved numerically. These calculations exhibit a discontinuity at the Hull cutoff magnetic field $B_H$ corresponding to the transition to magnetic insulation as observed analytically for a planar geometry. The ratio of the crossed-field SCLC to the nonmagnetic SCLC becomes essentially independent of geometry when we fix $\delta = \mathcal{D}/D_M > 0.6$, where $\mathcal{D}$ is the canonical gap distance accounting for geometric effects on electric potential and $D_M$ is the effective gap distance that accounts for magnetic field and geometry. The solutions for these geometries overlap as $\delta \to 1$ since the geometric corrections for electric potential and magnetic field match. This indicates the possibility of more generally accounting for the combination of geometric and magnetic effects when calculating $B_H$ and SCLC.

## I. INTRODUCTION

For numerous applications, including high-power microwaves, electric thrusters, and nano vacuum transistors, space-charge limited current (SCLC) is an important constraint that defines the maximum stable current for the system [1-3]. The seminal assessment of the SCLC in a one-dimensional (1D) planar diode in vacuum is given by the Child-Langmuir law (CLL), written as [1-6]

$$J_{CL} = \frac{4}{9}\epsilon_0\sqrt{\frac{2e}{m}}\frac{V^{3/2}}{D^2}, \tag{1}$$

where $e$ and $m$ are the electron charge and mass, respectively, $\epsilon_0$ is free space permittivity, $V$ is the electric potential drop across the diode, and $D$ is the gap distance. When the initial velocity of an electron is zero, the electric field at the cathode is zero for the SCLC; however, introducing a nonzero initial velocity shifts this zero electric field point into the gap to form a virtual cathode. This leads to two different solutions that have been referred to as the SCLC in the literature – one corresponds to electron reflection, which is more aptly called the "bifurcation" solution, while the second is the true SCLC [7].

Recently, the 1D SCLC has been derived for nonplanar geometries [8] using multiple techniques. One approach used variational calculus to extremize the current to obtain a coordinate system invariant solution for the SCLC [8-10]. We also applied variational calculus to extend the SCLC to nonzero initial velocity [11], which is relevant for thermionic cathodes [12], to recover both the bifurcation [13] and the true SCLC often attributed to Jaffé [14,15]. Another approach mapped the electric potential $\phi$ for a space-charge limited gap, which varies with the spatial position $x$ in 1D Cartesian coordinates as $\phi \propto x^{4/3}$, to the appropriate geometry of interest since such maps for $\phi$ are well known for many orthogonal geometries [10,16]. We have also applied Lie point symmetries [17], which are a subset of a broader category of techniques referred to as point transformations [18], to generalize both the CLL and recover the exact solution for the Jaffé solution for the SCLC with nonzero velocity [14,15]. Note that while variational calculus, conformal mapping, and point transformations are independent approaches for determining the SCLC, they all recover the *same* solutions for typical test cases (e.g., concentric cylinders, concentric spheres, tip-to-tip, and tip-to-plane).

These approaches are not limited to determining the SCLC in vacuum. We have recently demonstrated that they can be used to generalize the concept of nexus theory [19], which considers the linkages between SCLC and source emission mechanisms (e.g., field emission, thermal emission, and photoemission), to a tip-to-plane geometry in agreement with particle-in-cell simulations [20]. We have also demonstrated that we can apply point transformations to relativistic SCLC [18] and to gaps with collisionality that falls between vacuum (i.e., the CLL) and fully collisional (i.e., the Mott-Gurney law) [21]. These approaches may also be extended to multiple dimensions by leveraging variational calculus and vacuum capacitance [22].

An important expansion of these diodes is the application of a magnetic field perpendicular to the direction of the electric field across the gap. The crossed-field geometry is used for magnetrons [23], cyclotrons [23], nuclear fusion devices [24], and Hall thrusters [25]. Rather than traveling directly across the anode-cathode gap as in the absence of the magnetic field, electrons turn as they cross the gap due to the orthogonal magnetic field. This change in electron trajectory modifies the maximum current permissible in the gap. This behavior is further complicated since the strength of the magnetic field can have a significant impact on the electron trajectories.

For a sufficiently strong magnetic field in a 1D planar diode, an electron emitted from the cathode may be turned to such an extent that it cannot reach the anode. The Hull cutoff magnetic field $B_H$, which corresponds to the magnetic field at which an emitted electron reaches the anode with no velocity in the direction along the electric field, is given by [26]

$$B_H = \sqrt{\frac{2mV}{eD^2} + \left(\frac{mu_0}{eD}\right)^2}, \tag{2}$$

where $u_0$ is the initial velocity of an electron emitted from the cathode in the direction of the diode's electric field. Devices with $B > B_H$ are often referred to as magnetically insulated since emitted electrons do not cross the gap, meaning that there is no net current crosses the diode.

Assessing the maximum current permissible in crossed-field devices requires accurately accounting for this change of state. Lau *et al*. derived a condition for a magnetic analog to CL for $B < B_H$ [27]. Christenson *et al*. derived the classical SCLC condition based on zero electric field at the cathode [28] and a more complicated equation for the critical current density $J_c$ for a magnetically insulated diode ($B > B_H$) [29,30]. At $B_H$, $J_c$ decreases by a factor of two from $B \rightarrow B_H^-$ to $B \rightarrow B_H^+$. For $B > B_H$, electrons emitted from the cathode return to the anode in cycloidal orbits under stable conditions [30]. This cycloidal flow was sensitive to slight AC modulations [31] and small resistances representing circuit dissipation [32]. Slightly tilting the magnetic field $B$ along the electric field introduced a force in the direction of the electric field that eliminated magnetic insulation for any $B$ [33]. Electrons emitted under these conditions looped in the velocity-position phase space. The resulting space-charge buildup at the locations where the electrons changed direction dramatically reduced the critical current compared to the perfectly crossed-field geometry [33]. Magnetic insulation may also be impacted by the presence of ions or neutral particles, increasing transit times [34,35]. Considering gas in the gap as neutral particles that act as a friction force on the electrons eliminates magnetic insulation and causes electron trajectories to behave similarly to the tilted magnetic field [36]. As for the tilted magnetic field, collisions alter the limiting current in the crossed-field gap, leading to a crossed-field analog of the Mott-Gurney law [37]. A series resistor can also modify electron trajectories and magnetic insulation for $B \approx B_H$, which also modifies the limiting current [38].

While these studies demonstrate the ability to extend calculations of the limiting current of crossed-field diodes under 1D planar conditions, they provide limited utility for extending to more realistic geometries. Several studies have extended theories of electron trajectories and behavior in crossed-field geometries to more realistic geometries [39-45]. Others have used particle-in-cell simulations [46] and variational calculus [47] to determine the SCLC for a 1D cylindrical SCLC. However, none of these approaches, even the previous variational calculus approach, were readily extendable to theoretically describe the SCLC in a crossed-field gap in a general coordinate system.

In this paper, we modify the variational calculus approach used to determine the SCLC in a cylindrical crossed-field gap [47] to derive an equation that is valid for *any* orthogonal geometry and then obtain solutions for concentric spheres, tip-to-tip, and tip-to-plane geometries. Section II summarizes the derivation of the coordinate system invariant solution for both $B < B_H$ and $B > B_H$. Section III reports the application of these approaches to spherical, tip-to-tip, and tip-to-plane geometries, including the derivation of a scaling parameter that makes the SCLC relatively insensitive to geometry for a crossed-field diode. We make concluding remarks in Section IV.

## II. DERIVATION OF THE COORDINATE SYSTEM INVARIANT SOLUTION

### A. Governing Equations

Consider a diode with a grounded cathode at $x = C$ and an anode at $x = A$ held at a potential $V$ with a magnetic field $B$ applied in the $z$-direction, orthogonal to the electric field $E$ in the $x$-direction. Electrons are emitted from the cathode with initial velocity $u_0 = 0$ and current density $J$. Poisson's equation is given by

$$\nabla^2 \phi = \frac{\rho}{\epsilon_0} \tag{3}$$

with electric potential $\phi$, electric charge density $\rho$, and free space permittivity $\epsilon_0$. Electron continuity is given by

$$\vec{J} = \rho \vec{v}, \tag{4}$$

where $\vec{J}$ is the current density and $\vec{v}$ is the electron velocity in vector form. The nonrelativistic Lorentz force law is given by

$$m \frac{d\vec{v}}{dt} = e\left(\nabla\phi - \vec{v} \times \vec{B}\right), \tag{5}$$

where $t$ is time. Integrating (5) gives the total energy as

$$\frac{1}{2} m |\vec{v}|^2 = e\phi. \tag{6}$$

We can separate $\vec{v}$ into its components and solve by using conservation of momentum,

$$mn\left[\frac{\partial \vec{v}}{\partial t} + (\vec{v} \cdot \nabla)\vec{v}\right] = (-e)n\left(\vec{E} + \vec{v} \times \vec{B}\right) - \nabla p, \tag{7}$$

where $n$ is the number of particles and $p$ is pressure. Assuming cold beam injection (thus $\nabla p = 0$) and steady state (all time derivatives are zero) simplifies (7) to

$$m[(\vec{v} \cdot \nabla)\vec{v}] = (-e)\big(\vec{E} + \vec{v} \times \vec{B}\big). \tag{8}$$

For a fixed voltage, the current $I$ with respect to the current density $J$ is

$$I = \int JdS\,, \tag{9}$$

where $dS$ is the differential area at the location where $J$ is calculated (e.g., the cathode when considering the SCLC).

Applying this process to different geometries requires solving this problem in general form rather than for a specific coordinate system (e.g., Cartesian coordinates, as described above). An effective way to accomplish this is by defining scale factors $h_1, h_2$, and $h_3$, as we have done previously for variational calculus [9] and point transformations [17,18]. We can write the 1D Laplacian using scale factors as

$$\nabla^2 \phi = \frac{1}{h_1 h_2 h_3} \frac{\partial}{\partial q_1} \left[ \frac{h_2 h_3}{h_1} \frac{\partial \phi}{\partial q_1} \right] \tag{10}$$

and the material derivative as

$$(\vec{v} \cdot \nabla)\vec{v} = \left[ \frac{v_1 v_1'}{h_1} - \frac{v_2^2}{h_1 h_2} \frac{\partial h_2}{\partial q_1} \right] \hat{q}_1 + \left[ \frac{v_1 v_2'}{h_1} + \frac{v_1 v_2}{h_1 h_2} \frac{\partial h_2}{\partial q_1} \right] \hat{q}_2, \tag{11}$$

where the subscripts 1 and 2 refer to the vector components in the directions of the electric field and orthogonal to both the electric and magnetic fields, respectively; $q, v$, and $h$ are the spatial variable, velocity, and scale factor, respectively; and the prime ($'$) denotes differentiation with respect to $q_1$. Combining (8) and (11) gives one equation in the $\hat{q}_1$ direction and a second in the $\hat{q}_2$ direction. The $\hat{q}_1$ equation provides the same information as energy conservation, so it is redundant. The simplified $\hat{q}_2$ equation in terms of the scale factors and the velocity components is given by

$$\frac{v_2'}{h_1}+\frac{v_2}{h_1 h_2}\frac{\partial h_2}{\partial q}=\Omega, \tag{12}$$

where $\Omega = eB/m$ is the electron cyclotron frequency and we set $q_1 = q$ since no future equations need more than just the first spatial variable. This ordinary differential equation can be solved for $v_2$ using the boundary condition $v_2(C) = 0$, where $C$ represents the cathode position, to obtain

$$v_2=\frac{\Omega}{h_2}\int_C^q h_1 h_2 dq\,. \tag{13}$$

Substituting (13) into energy conservation (6) gives $v_1$ as a function of $\phi$ as

$$v_1=\sqrt{\frac{2e\phi}{m}-\frac{\Omega^2}{h_2^2}\left(\int_C^q h_1 h_2 dq\right)^2}\,. \tag{14}$$

To determine the maximum permissible current, we appeal to the second order Euler-Lagrange equation with no friction factors, given by [48]

$$\frac{\partial^2}{\partial q^2}\frac{\partial f}{\partial \phi_{qq}}-\frac{\partial}{\partial q}\frac{\partial f}{\partial \phi_q}+\frac{\partial f}{\partial \phi}=0, \tag{15}$$

where the subscript $q$ represents the derivative of $\phi$ with respect to $q$ and $f$ represents the equation inside the integral for the current $I$ in the crossed-field gap. Applying the Euler-Lagrange equation requires an integral between two distinct points for extremization. Combining (3), (4), (9), (10), and (14) gives the current density evaluated at the cathode as

$$J=\varepsilon_0\frac{\partial}{\partial q_1}\left[\frac{h_2 h_3}{h_1}\frac{\partial \phi}{\partial q_1}\right]\sqrt{\frac{2e\phi}{m}-\frac{\Omega^2}{h_2^2}\left(\int_C^q h_1 h_2 dq\right)^2}\,. \tag{16}$$

The resulting current

$$I = \int \varepsilon_0 \frac{\partial}{\partial q}\left[\frac{h_2 h_3}{h_1}\frac{\partial \phi}{\partial q}\right]\sqrt{\frac{2e\phi}{m} - \frac{\Omega^2}{h_2^2}\left(\int_C^q h_1 h_2 dq\right)^2}\, ds = \int f dSs\,, \tag{17}$$

where $dS$ is the surface area of the cathode. Equation (17) satisfies the requirements of variational calculus, as it is an integral over distinct bounds, where the integrand $f$ is twice continuously differentiable with respect to $q$, $\phi$, and $\phi'$, and $\phi$ is also twice continuously differentiable with respect to $q$. Therefore, $f$ can be substituted directly into (15). Note that (17) is *not*, in general, the critical current $I_C$; obtaining $I_C$ requires first extremizing $\phi$ using variational calculus. Thus, (16) and (17) become $J_C$ and $I_C$, respectively, upon extremizing $\phi$ using variational calculus.

As described above, the vacuum crossed-field condition requires different solutions for $B < B_H$ and $B > B_H$ since a discontinuity in the limiting current occurs at $B = B_H$ [30]. For a planar diode, $B_H$ is given by (2). We can derive $B_H$ in general coordinates by solving (14) for $v_1(q = A) = 0$ to obtain

$$B_H = \sqrt{\frac{2mV}{e}}\left|\frac{h_2|_{q=A}}{\int_C^A h_1 h_2 dq}\right| = \sqrt{\frac{2mV}{e}}\frac{1}{|D_M|}, \tag{18}$$

where $D_M$ is an effective gap distance that represents the geometric influence on the magnetic field. Equation (18) holds for any orthogonal geometry given Newtonian physics. For the cases studied here with negligible initial velocity and gap voltages on the order of Volts and kilovolts, there will be no contributing relativistic effects. Table I summarizes $D_M$ and other gap distance scalings that we shall introduce in this study. For a planar geometry, $D_M = D$, making (18) identical to (2) with $u_0 = 0$. Poisson's equation (3) and continuity (4) are the primary equations for solving for SCLC. Extending this approach to canonical coordinates requires combining the Euler-Lagrange equation from (15) with the equation for the extremized current in a crossed-field gap from (17). The result is in canonical coordinates [17, 18] and may be converted to the desired coordinate system by applying the appropriate scale factors for the geometry of interest.

TABLE I: Definitions of gap distance scalings used throughout the manuscript, where $D_M$ adjusts the Hull cutoff magnetic field for nonplanar geometries, $D_a$ eliminates the angular dependence in the tip-to-tip and tip-to-plane geometries, and $\mathcal{D}$ accounts for changes in the electric potential in nonplanar geometries without a magnetic field.

| | $D_M$ | $D_a$ | $\mathcal{D}$ |
|---|---|---|---|
| Magnetic Field Scaling | $B_H \sqrt{\frac{2mV}{e}} \frac{1}{\lvert D_M \rvert}$ | $B_a = \sqrt{\frac{2mV}{e}} \frac{1}{\lvert D_a \rvert}$ | $B_{\mathcal{D}} = \sqrt{\frac{2mV}{e}} \frac{1}{\lvert \mathcal{D} \rvert}$ |
| Planar | $A - C$ | $A - C$ | $A - C$ |
| Cylindrical | $\frac{R_A}{2} \lvert 1 - a_r^2 \rvert$ | – | $R_C \ln(R_A/R_c)$ |
| Spherical | $\frac{R_A}{2} \lvert 1 - a_r^2 \rvert$ | – | $R_C(R_A - R_C)/R_A$ |
| Tip-to-Tip | $\frac{a(A - C - \cos(A)\sin(A) + \cos(C)\sin(C))}{2\sin(A)}$ | $a$ | $a\sin^2(C) \ln\left[\frac{\tan(A/2)}{\tan(C/2)}\right]$ |
| Tip-to-Plane | $\frac{a[\pi - 2C + \sin(2C)]}{4}$ | $a$ | $a\sin^2(C)\ln[\cot(C/2)]$ |

**B. Coordinate system invariant solution for $B < B_H$**

Evaluating the integrand $f$ from (17) in (15) provides an additional equation completely describing the electric potential for $B < B_H$ and the electric potential in the region containing space-charge for $B > B_H$, as shown in the Appendix. Applying (A2) to the general form of the Laplacian (1) and simplifying gives

$$
\begin{aligned}
\nabla^2\phi &= \left(\frac{h_2'}{h_2 h_1^2} + \frac{h_3'}{h_3 h_1^2} - \frac{h_1'}{h_1^3}\right)\phi' + \frac{\phi''}{h_1^2} \\
&= \frac{2B^2 e m h_1^3 h_2^5 \phi + B^2 e\left(\int_C^q h_1 h_2 dq\right)^2 \left(2m h_2^2\,\phi - B^2 e\left(\int_C^q h_1 h_2 dq\right)^2\right) h_1' h_2'}{2m h_1^3 h_2^3 \left(2m h_2^2 \phi - B^2 e\left(\int_C^q h_1 h_2 dq\right)^2\right)} \\
&- \frac{B^2 e \int_C^q h_1 h_2 dq \left(4m h_2 h_3 \phi h_2' + B^2 e\left(\int_C^q h_1 h_2 dq\right)^2 h_3' + 2m h_2^2 (h_3 \phi' - \phi h_3')\right)}{2m h_1 h_2 h_3 \left(2m h_2^2 \phi - B^2 e\left(\int_C^q h_1 h_2 dq\right)^2\right)} \\
&+ \frac{4B^2 e m h_2^2 \phi\left(\int_C^q h_1 h_2 dq\right)^2 h_2'^2 - B^4 e^2\left(\int_C^q h_1 h_2 dq\right)^4 h_2'^2 + m^2 h_2^6 \phi'^2}{2m h_1^2 h_2^2 h_2^4 \left(2m h_2^2 \phi - B^2 e\left(\int_C^q h_1 h_2 dq\right)^2\right)} \\
&+ \frac{B^4 e^2 h_2\left(\int_C^q h_1 h_2 dq\right)^4 (h_2' h_3' + h_3 h_2'') - 2B^2 e m h_2^3\left(\int_C^q h_1 h_2 dq\right)^2 \left(h_2'(\phi h_3' - h_3 \phi')\right) + h_3 \phi h_2''}{2m h_1^2 h_2^2 h_2^4 h_3 \left(2m h_2^2 \phi - B^2 e\left(\int_C^q h_1 h_2 dq\right)^2\right)},
\end{aligned} \tag{19}
$$

which holds for *any* orthogonal coordinate system. Equation (19) assumes no relativistic effects, zero initial velocity, and no collisions. Applying (19) to a real diode requires choosing or approximating the diode such in an orthogonal coordinate system. Generally, the cathode and anode will be in fixed locations in one coordinate and infinite in the other two coordinates (tip-to-tip and tip-to-plate differ by assuming that one coordinate must go to zero to achieve 1D, meaning that they are evaluated at a single point) with the magnetic field in one of the infinite directions to be perpendicular to the electric field in the finite coordinate direction. The critical current can be obtained from (19) as long as all three coordinates (directions) remain orthogonal.

### C. Coordinate system invariant solution for $B > B_H$

For $B > B_H$, the electrons do not reach the anode, creating a region in the diode without space-charge. Including the region without space-charge requires writing an equation for the electric potential across the gap as

$$
\phi(q) = \frac{B\Omega}{2h_2^2}\left(\int_C^q h_1 h_2 dq\right)^2. \tag{20}
$$

The electric field $E \equiv -\nabla\phi$ within the space-charge region is given by

$$E_S = \frac{B\Omega \int_C^q h_1 h_2 dq \left(\int_C^q h_1 h_2 dq\, h_2' - h_1 h_2^2\right)}{h_1 h_2^3}. \tag{21}$$

For the region without space-charge, the electric field is continuous, so $\nabla \cdot E_V = 0$ [47]. At the boundary between the space-charge and vacuum (i.e., no electrons) regions, $E_S(q = H) = E_V(H < q \leq A)$, where $H$ represents the hub height, or the maximum excursion of the electron across the gap, and $E_V$ represents the constant electric field within the vacuum region. This gives

$$E_V = \frac{1}{h_2 h_3}\left[h_2 h_3 \frac{B\Omega \int_C^q h_1 h_2 dq \left(\int_C^q h_1 h_2 dq\, h_2' - h_1 h_2^2\right)}{h_1 h_2^3}\right]\Bigg|_{q\to H}. \tag{22}$$

Integrating (21) across the space-charge region from the cathode $C$ to the hub height $H$ and (22) across the vacuum region from the hub height $H$ to the anode $A$ gives the total potential drop across the gap (i.e., anode voltage $V$) as

$$\begin{aligned} V = &\int_C^H \frac{eB^2 \int_C^q h_1 h_2 dq \left(\int_C^q h_1 h_2 dq\, h_2' - h_1 h_2^2\right)}{m h_1 h_2^3} dq \\ &+ \left[h_2 h_3 \frac{eB^2 \int_C^q h_1 h_2 dq \left(\int_C^q h_1 h_2 dq\, h_2' - h_1 h_2^2\right)}{m h_1 h_2^3}\right]\Bigg|_{q\to H} \int_H^A \frac{1}{h_2 h_3}\, dq. \end{aligned} \tag{23}$$

Numerically solving (23) gives the hub height $H$, which provides the final boundary condition for obtaining the limiting current for $B > B_H$ assuming a nonrelativistic geometry with no collisions and negligible initial velocity.

## III. APPLICATION OF THE COORDINATE SYSTEM INVARIANT SOLUTION TO SAMPLE GEOMETRIES

We now apply these general solutions to specific geometries, starting with confirming the prior solutions [47] for the SCLC for planar and cylindrical crossed-field geometries before deriving solutions for spheres, tip-to-tip, and tip-to-plane geometries.

### A. Planar Geometry

We first verify that we can recover the planar solution [47]. Figure 1 shows the coordinate system and projected electron paths for $B = 0, B < B_H$, and $B > B_H$. The scale factors for planar or Cartesian coordinates are unity (i.e., $h_1 = h_2 = h_3 = 1$). Combining these with (19) yields

$$\nabla^2\phi = \phi'' = \frac{2B^2e\phi^2 - 2B^2e(x-C)\phi' + m\phi'^2}{2(2m\phi - B^2e(x-C)^2)}, \tag{24}$$

where the prime ($'$) denotes differentiation with respect to $x$. Combining (24) with $\Omega = eB/m$ and setting the cathode at $C = 0$ gives

$$\nabla^2\phi = \phi'' = \frac{\phi'^2 - 2B\phi' x + 2B\Omega\phi}{4\phi - 2B\Omega x^2}. \tag{25}$$

Equation (25) exactly matches the related solution from applying the planar Laplacian using variational calculus [47], which means that it will recover the same SCLC.

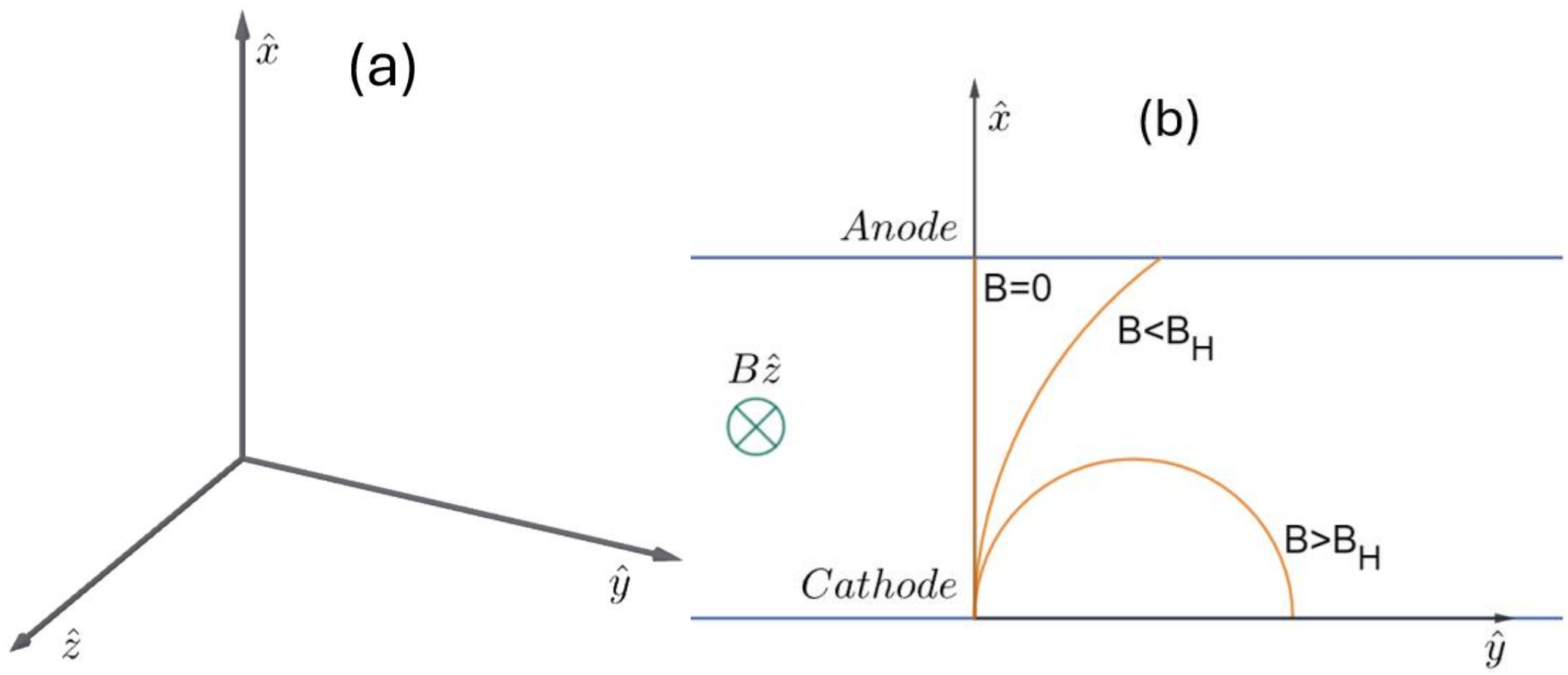

**FIG. 1.** Geometry for a planar diode. (a) The Cartesian coordinate system used for a planar diode. (b) Schematic demonstrating different electron trajectories in orange for $B = 0$, $B < B_H$, and $B > B_H$.

### B. Cylindrical Geometry

We next demonstrate that the approach outlined in Sec. II recovers the solution for the SCLC in a cylindrical crossed-field diode [47]. Figure 2 shows the coordinate system and projected electron paths for a crossed-field diode with $B < B_H$ and $B > B_H$. Applying the cylindrical scale factors, $h_1 = h_3 = 1$ and $h_2 = r$, to (19) yields

$$\nabla^2\phi = \frac{\phi'}{r} + \phi''$$

$$= \frac{-B^4e^2\left(\frac{r^2}{2}-\frac{R_C^2}{2}\right)^4 + 2B^2emr^6\phi + 4B^2emr^2\left(\frac{r^2}{2}-\frac{R_C^2}{2}\right)^2\phi + 2B^2emr^3\left(\frac{r^2}{2}-\frac{R_C^2}{2}\right)^2\phi'}{2mr^4\left[2mr^2\phi - B^2e\left(\frac{r^2}{2}-\frac{R_C^2}{2}\right)^2\right]}$$

$$+\frac{m^2r^6\phi'^2 - B^2er^3\left(\frac{r^2}{2}-\frac{R_C^2}{2}\right)(4mr\phi + 2mr^2\phi')}{2mr^4\left[2mr^2\phi - B^2e\left(\frac{r^2}{2}-\frac{R_C^2}{2}\right)^2\right]}, \tag{26}$$

where $R_C$ is the cathode radius and the prime ($'$) denotes differentiation with respect to $r$. Defining $\Omega = eB/m$ reduces (26) to

$$\nabla^2\phi = \frac{B^2\Omega^2(R_C^2 - r^2)^4 - 16B\Omega r^2(r^4 + R_C^4)\phi + 8B\Omega r^3(r^4 - R_C^4)\phi' - 16r^6\phi'^2}{8r^4[B\Omega(r^2 - R_C^2)^2 - 8r^2\phi]}. \qquad (27)$$

Multiplying the numerator and denominator of (27) by $-16^{-1}r^{-6}$ yields

$$\nabla^2\phi = \frac{\phi'^2 - \frac{r\phi' B\Omega}{2}\left(1 - \left(\frac{R_C}{r}\right)^4\right) + B\Omega\phi\left(1 + \left(\frac{R_C}{r}\right)^4\right) - \frac{B^2\Omega^2 r^2}{16}\left(1 - \left(\frac{R_C}{r}\right)^2\right)^4}{4\phi - \frac{B\Omega r^2}{2}\left(1 - \left(\frac{R_C}{r}\right)^2\right)^2}, \qquad (28)$$

which exactly matches the previous derivation using variational calculus [47]. We can also verify the Hull cutoff magnetic field for this geometry by applying the cylindrical scale factors to (19) and defining $R_A$ as the anode radius to obtain

$$B_{H,cyl} = \sqrt{\frac{2mV}{e}}\left|\frac{2R_a}{R_A^2 - R_c^2}\right| = \sqrt{\frac{2mV}{e}}\frac{1}{\frac{R_A}{2}\left|1 - \frac{R_c^2}{R_A^2}\right|}, \qquad (29)$$

which matches Ref. [47].

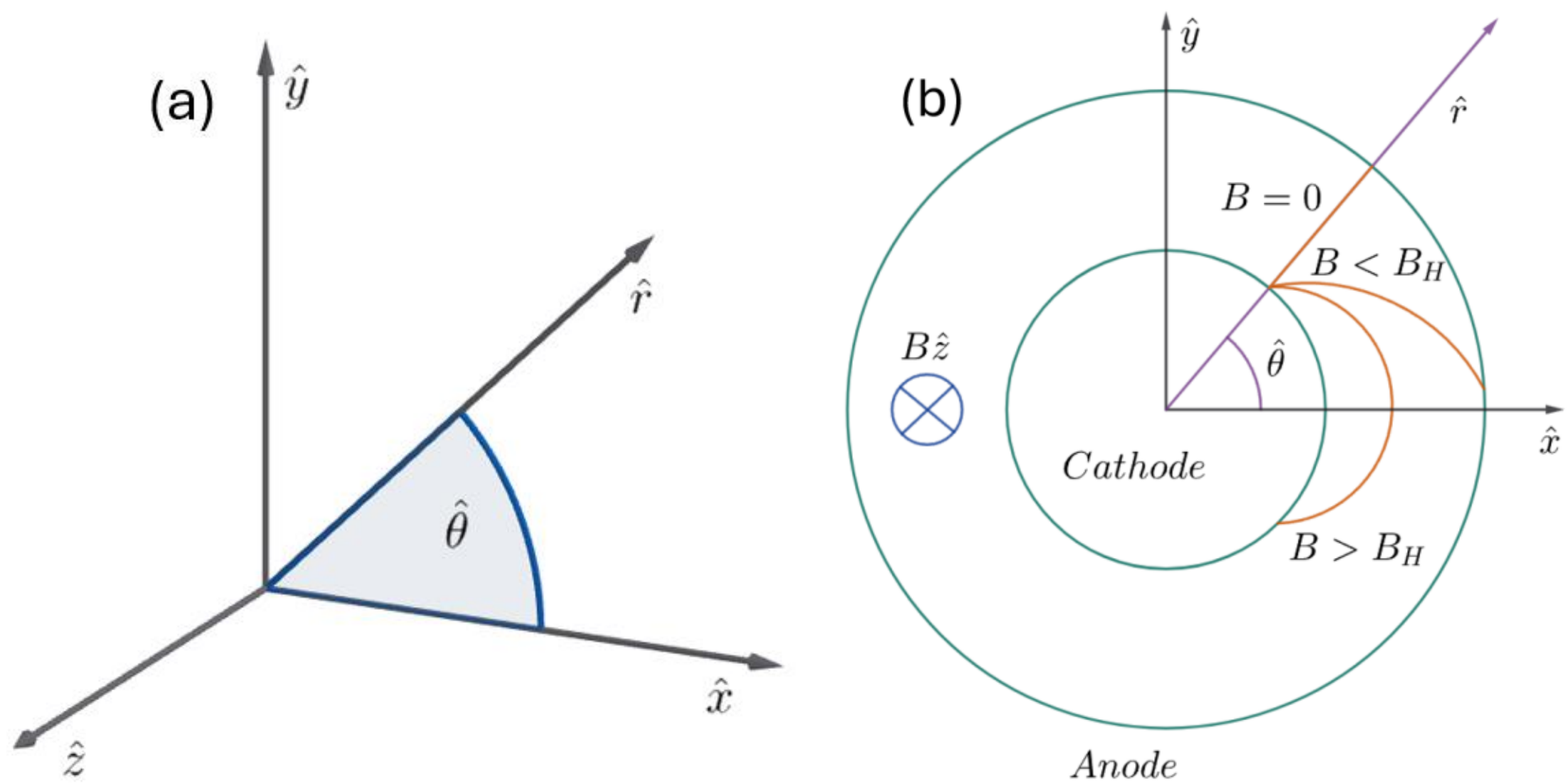

**FIG. 2.** Geometry for a cylindrical diode. (a) The cylindrical coordinate system. (b) Cross-section of a cylindrical diode with magnetic field $B\hat{z}$ demonstrating different electron trajectories for $B = 0$, $B < B_H$, and $B > B_H$.

### C. Spherical Geometry

For a spherical geometry, the three coordinates are $r, \theta,$ and $\psi$. Assuming no electron-to-electron interactions and a vacuum diode means that there will be no electron flow in the $\varphi$-direction. Figure 3 the coordinate system and cross-section for a spherical crossed-field diode. While the cylindrical [cf., Fig. 2(b)] and spherical [cf., Fig. 3(b)] have the same cross-section, the magnetic fields are different since $\vec{B} = B\hat{z}$ for cylindrical and $\vec{B} = B\hat{\psi}$ in spherical.

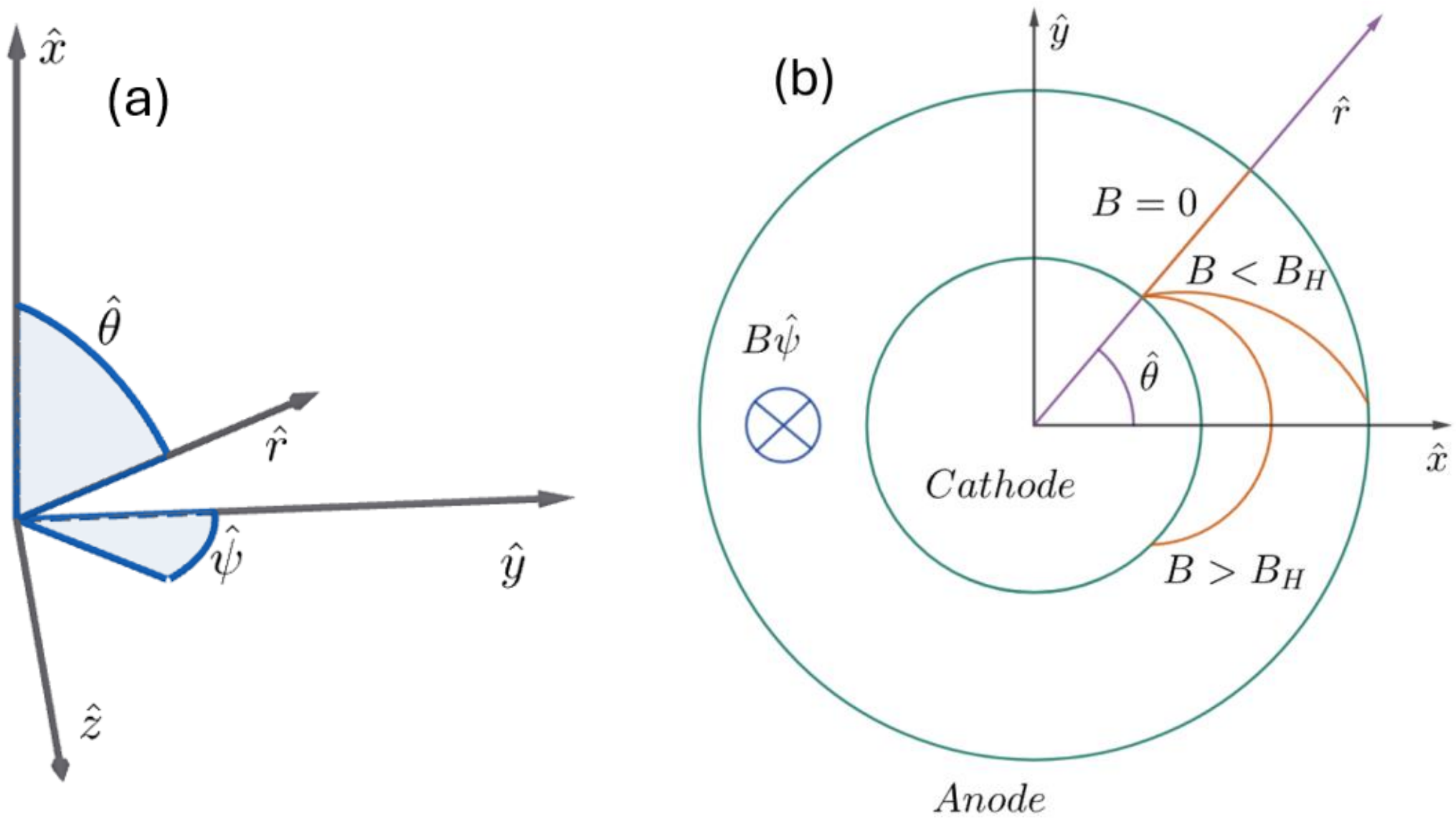

**FIG. 3.** Geometry for a spherical diode. (a) The spherical coordinate system. (b) Cross-section of a spherical diode with magnetic field $B\hat{\psi}$ demonstrating different electron trajectories for $B = 0$, $B < B_H$, and $B > B_H$. Note that although the two-dimensional cross-section as the cylindrical case [cf., Fig. (2b)], the magnetic field direction and overall three-dimensional geometries differ significantly.

Due to path symmetries, there will also be no $\theta$-dependence, making our 1D spherical behavior only a function of $r$ and $B$. This allows us to define the 1D scale factors for spherical coordinates as $h_1 = 1$ and $h_2 = h_3 = r$. Substituting these scale factors into (19) for $B < B_H$ yields

$$\nabla^2\phi = \frac{B^4e^2(R_C^2 - r^2)^3 + 4B^2em(R_C^4 + 3r^4)\phi + 4B^2emr(R_C^4 - r^4)\phi' + 8m^2r^4\phi'^2}{4mr^2[8mr^2\phi - B^2e(R_C^2 - r^2)^2]}, \tag{30}$$

where the prime ($'$) denotes differentiation with respect to $r$. Nondimensionalizing (30) by defining $a_r = R_C/R_A$, $\bar{r} = r/R_A$, $\bar{\phi} = \phi/V$, and $\bar{B} = B/B_{H,sph}$, where $B_{H,sph}$ is the Hull cutoff for a spherical diode, found from (18) as

$$B_{H,sph} = \sqrt{\frac{8mV}{e}} \left| \frac{R_A}{R_A^2 - R_C^2} \right| = \sqrt{\frac{2mV}{e}} \frac{1}{\frac{R_A}{2} |1 - a_r^2|}, \tag{31}$$

which has the same form as $B_{H,cyl}$, gives

$$\nabla^2 \bar{\phi} = \frac{(1-a_r^2)^4 \bar{r}^4 \bar{\phi}'^2 + 4(1-a_r^2)^2 \bar{B}^2 \bar{r}(a_r^4 - \bar{r}^4)\bar{\phi}' + 4(1-a_r^2)^2 \bar{B}^2 (a_r^4 + 3\bar{r}^4)\bar{\phi} - 8\bar{B}^4(\bar{r}^2 - a_r^2)^3}{4(1-a_r^2)^2 \bar{r}^2 [(1-a_r^2)^2 \bar{r}^2 \bar{\phi} - \bar{B}^2 (a_r^2 - \bar{r}^2)^2]}, \tag{32}$$

where the prime ($'$) denotes differentiation with respect to $\bar{r}$. Substituting the numerical solution of (32) with the boundary conditions $\bar{\phi}(0) = 0$ and $\bar{\phi}'(0) = 0$ into (3) and (4) gives the SCLC $J_C$. The nondimensionalization makes (32) a function of only $a_r$ and $\bar{B}$.

Evaluating SCLC for $B > B_H$ requires evaluating (23) to solve for the hub height $H$ and then evaluating the derivative of electric potential with respect to $\bar{r}$ from (32) at the hub height. In spherical coordinates, substituting the scale factors into (23) and nondimensionalizing as immediately above with $\bar{H} = H/R_A$ gives a nondimensionalized version of (23) as

$$1 = \frac{\bar{B}^2[\bar{H}^4(3 - 2\bar{H}) - 2a_r^2 \bar{H}^2 + a_r^4(2\bar{H} - 1)]}{(a_r^2 - 1)\bar{H}^2} \tag{33}$$

and a nondimensionalized version of (20) as

$$\bar{\phi}(\bar{r}) = \frac{\bar{B}^2(a_r^2 - \bar{r}^2)}{\bar{r}^2(a_r^2 - 1)^2}. \tag{34}$$

The derivative of (34) evaluated at the hub provides the additional equation to solve for the SCLC for $B > B_H$ as

$$\bar{\phi}'(\bar{H}) = \frac{2\bar{B}^2(\bar{H}^4 - a_r^4)}{\bar{H}^3(a_r^2 - 1)^2}. \tag{35}$$

Using (35) as a boundary condition for the numerical solution allows us to extend the calculation to $B > B_H$.

We next define the SCLC for a spherical geometry without a magnetic field as [9, 17, 22]

$$J_{SCLC} = \frac{4V^{3/2}\epsilon_0\sqrt{2e/m}}{9a_r^2|R_A - R_C|^2}. \tag{36}$$

This allows us to scale the SCLC for the spherical crossed-field diode directly to the SCLC for a 1D spherical diode, analogous to the previous planar scalings with the CLL and the Jaffé solution. Figure 4 shows $\bar{J} = J_C/J_{SCLC}$ as a function of $\bar{B} = B/B_H$ for a spherical crossed-field diode, as well as the previous planar solution for comparison. As the magnetic field vanishes ($\bar{B} \to 0$), $\bar{J} \to 1$, indicating the expected behavior of approaching the 1D spherical solution. The numerical approaches used limit how close we can approach $\bar{B} \to 1$; thus, all calculations for $B > B_H$ started at $B \approx 1.1B_H$, which caused us to miss some of the trends as $\bar{B} \to 1^+$, such as the nonmonotonic spike near $B_H$. The spherical solution exhibits similar trends to the planar [30] and cylindrical solutions [47], with a monotonic change with $a$ for both small and large $\bar{B}$. The primary difference is that decreasing $a_r$ reduces $J_C/J_{SCLC}$ as $\bar{B} \to 1^-$ but increases $J_C/J_{SCLC}$ as $\bar{B} \to 1^+$ to the left of the local minimum.

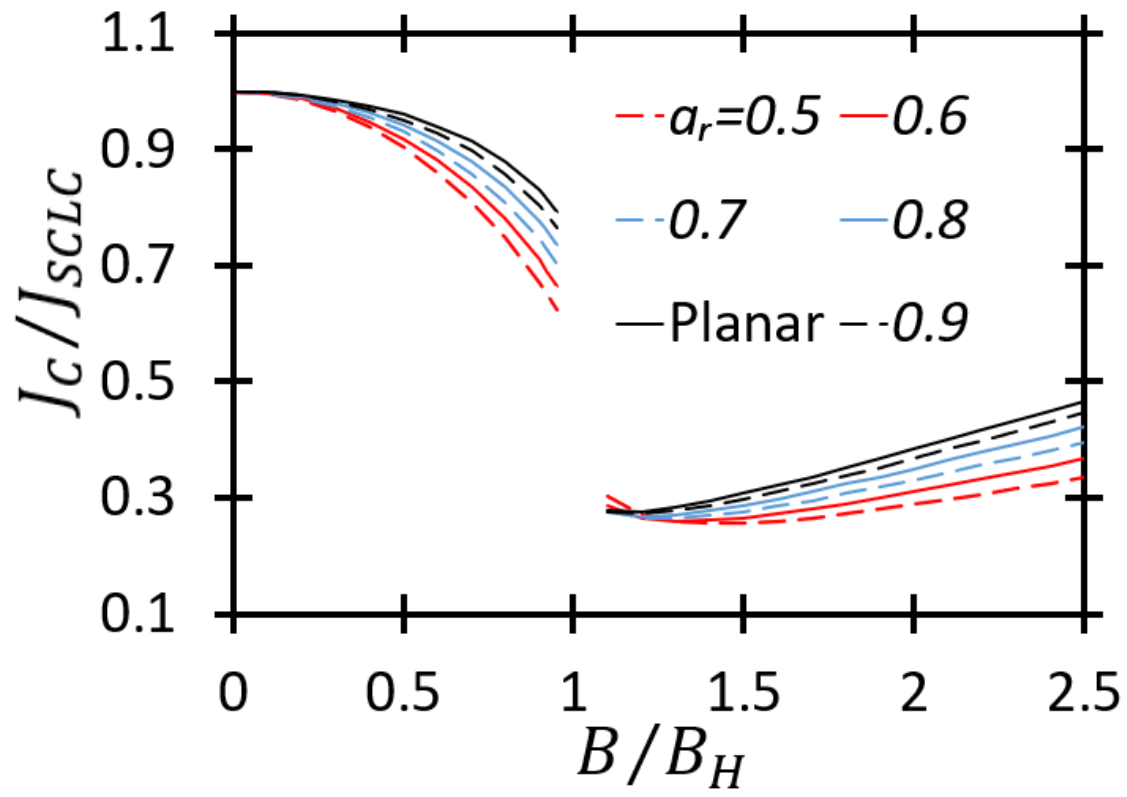

**FIG. 4.** Space-charge-limited current (SCLC) for a spherical crossed-field diode $J_C$ normalized to the SCLC for a nonmagnetic spherical diode $J_{SCLC}$ (36) as a function of magnetic field $B$ normalized to the spherical Hull cutoff magnetic field $B_H$ (31) for various $a_r = R_C/R_A$, where $R_C$ is the cathode radius and $R_A$ is the anode radius. The planar result [27-30] is included for comparison.

**D. Tip-to-tip geometry**

The general approach outlined above also applies to tip-to-tip and tip-to-plane geometries; however, there is no standard sharp tip geometry. This has been addressed previously by using spheroidal coordinates [10]. For the spheroidal coordinate system, a rounded tip is formed by creating lines orthogonal to a surface circle, as shown in Fig. 5. Using the angle $\vartheta$ of this tip as the 1D parameter with $\mu = 0$ creates sharp tips using spheroidal coordinates. We use a prolate spheroidal geometry to closely approximate the sharpest possible tip to best represent real-life tips. In this geometry, $\vartheta$ represents the angle of this tip. Additionally, we define the cathode angle $\vartheta_C = C$ and the anode angle $\vartheta_A = A$. In this geometry, we assume that the magnetic field runs in the angular direction, so it would be spheroidal in shape but would not have any nonphysical conditions as the tips would be in the way of the magnetic fields in these locations.

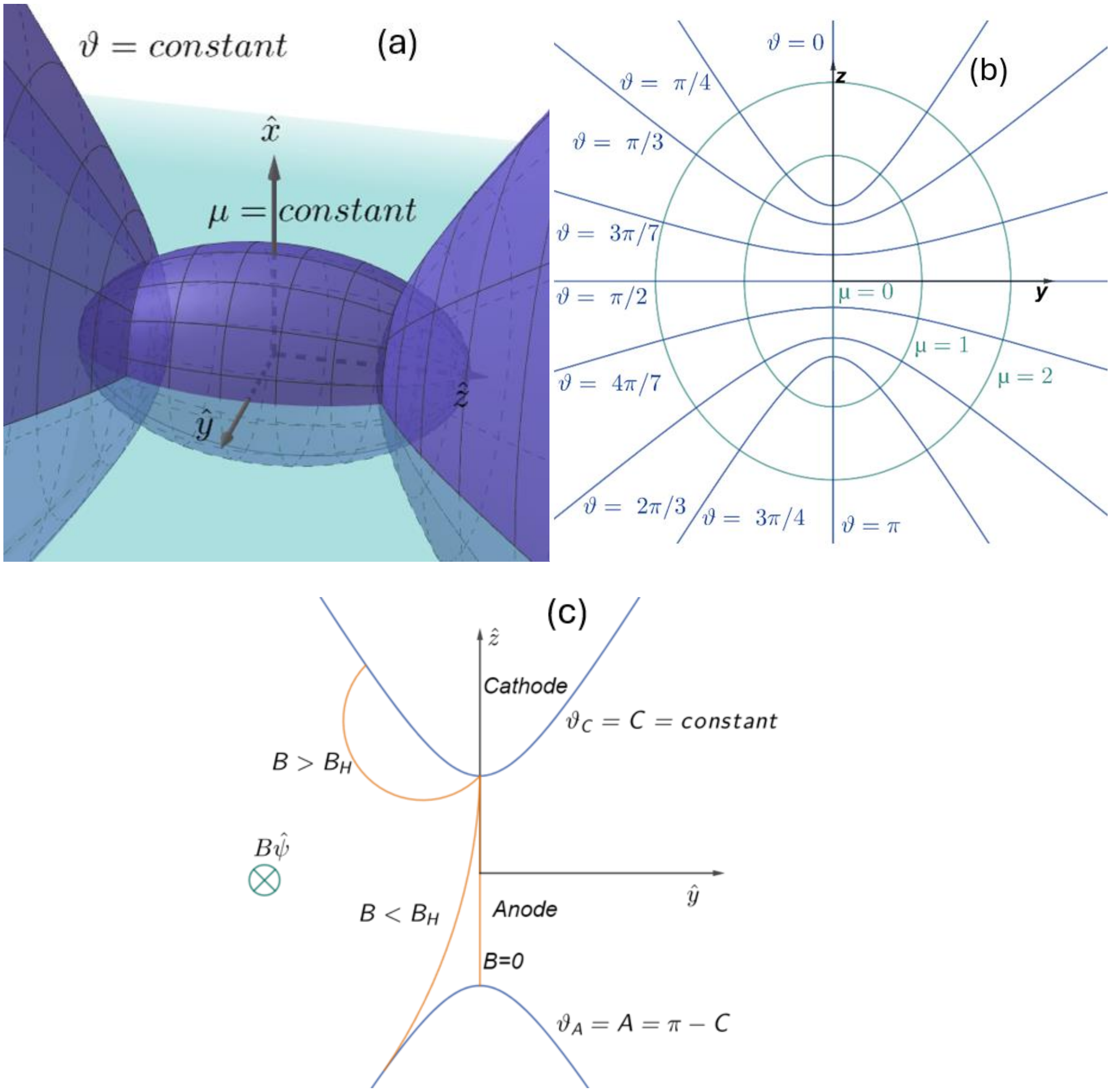

**FIG. 5.** Prolate spheroidal coordinate system used for the tip-to-tip geometry. (a) Three-dimensional (3D) geometry for a constant angle $\vartheta$ and length $\mu$. The distance between the foci and origin $a$ is unity. The $q$ variable in this geometry is $\vartheta$. (b) 2D perspective of the tips in the $yz$-plane in Cartesian coordinates, which can be rotated about the $z$-axis to produce the 3D geometry from (a). (c) Representative diode showing estimated electron trajectories for $B = 0$, $B < B_H$, and $B > B_H$.

We obtain the equations for a spheroidal geometry by substituting the 1D scale factors, $h_1 = h_2 = h_3 = a\sin(\vartheta)$, where $a$ is the distance between the foci and the origin, into the general form (19) to obtain (B3) for $\nabla^2\phi$.

Note that we do not consider variations in the physical distance $a$ in this analysis (and our subsequent analysis for the tip-to-plane geometry), which we set to unity; we only consider the variation of the angle $\vartheta$, which has units of radians and defines the sharpness of the tip. Since the nondimensionalization used in this geometry cancels out the physical distance $a$ as both the left and right-hand sides of (B3) are functions of $1/a^2$, we can set $a$ to any nonzero value without loss of generality. Numerically solving (B3) requires setting the boundary conditions as $\phi(0) = 0$ and $\phi(A) = V$, which become $\bar{\phi}(C) = 0$ and $\bar{\phi}(A) = 1$, respectively, for $B < B_H$ with boundary conditions $\bar{\phi}'(C) = 0$ and the derivative of the potential from (20) with respect to $\vartheta$ at the hub angle $\vartheta = H$ to obtain

$$\bar{\phi}'(H) = \frac{2\bar{B}^2 \sin(A)(\cos(H)+\csc(H)(C-H-\cos(C)\sin(C))}{(A-C-\cos(A)\sin(A)+\cos(C)\sin(C))^2} \times \{-\csc(H) - \cot(H)\csc(H)\,[C - H - \cos(C)\sin(C)] - \sin(H)\}. \tag{37}$$

We find $H$ by applying the scale factors to (23) and nondimensionalizing as above to obtain

$$1 = \frac{2\bar{B}^2 \sin^2(A)\sin^2(H)\ln(\cot(H/2)\tan(A/2))\,\{\cos(H) + \csc(H)\,[C - H - \cos(C)\sin(C)]\}}{[A - C - \cos(A)\sin(A) + \cos(C)\sin(C)]^2 - \frac{\csc^4(H)\,(2C - 2H - \sin(2C) + \sin(2H))^2}{[2A - 2C - \sin(2A) + \sin(2C)]^2}}, \tag{38}$$

which can be solved numerically.

Figure 6 compares the SCLC for crossed-field diodes with tip-to-tip geometries for various $\vartheta$ and $a = 1$ scaled to the SCLC for a tip-to-tip geometry without a magnetic field, given by [9, 17, 20]

$$J_{SCLC} = \frac{4V^{3/2}\epsilon_0\sqrt{2e/m}}{9\{a\sin^2(C)\ln[\tan(A/2)/\tan(C/2)]\}^2}. \tag{39}$$

Figure 6 is a representation of parallel tips, where $A = \pi - C$ and C varies. For $B < B_H$ and $B \gg B_H$, $\bar{J}$ diverges from the planar result with decreasing $C$. For $B_H < B \lesssim 1.6B_H$, $\bar{J}$ converges to planar with increasing tip angles; however, the sharpest tip studied, $C = \pi/6$, behaves noticeably different with increasing $B$ throughout this range.

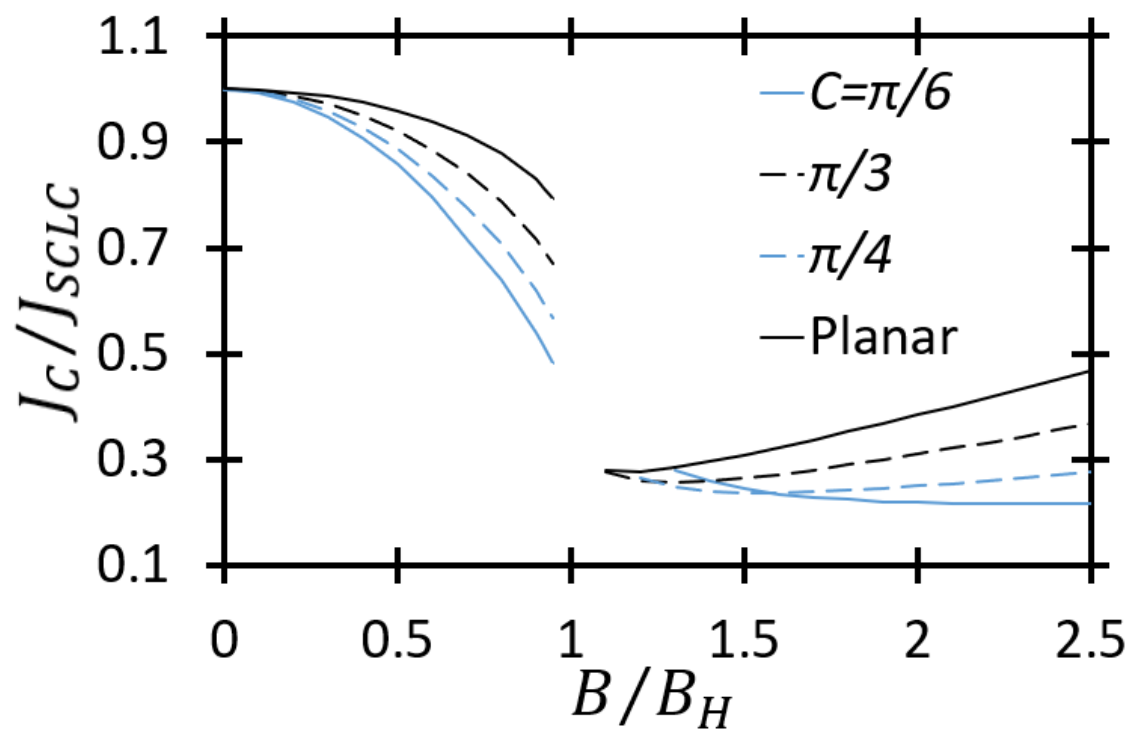

**FIG. 6.** Space-charge-limited current (SCLC) for a tip-to-tip crossed-field diode $J_C$ normalized to the SCLC for a nonmagnetic tip-to-tip diode $J_{SCLC}$ (39) as a function of magnetic field $B$ normalized to the Hull cutoff magnetic field $B_H$ (B2) for tips of various $\vartheta$ compared to the planar result [27-30].

Near the Hull cutoff, the curves align less with increasing tip sharpness. It becomes impossible to numerically solve for the SCLC for $B \approx B_H$. To further explore this behavior, we rescaled the magnetic field $B$ to an angle-independent variable called $B_a$, given by

$$B_a = \sqrt{\frac{2mV}{eD_a^2}}, \tag{40}$$

which becomes the planar Hull cutoff magnetic field for planar diodes when $D_a = D$. For tip-to-tip (and our subsequent tip-to-plane) geometries, removing the angle effects on the magnetic field axis requires setting $D_a = a$. This provides the additional benefit of eliminating the angle-dependence in the direction of the electric field to further elucidate how the angles influenced the SCLC by scaling to $B_a$ instead of $B_H$.

Figure 7 shows how scaling $B$ to $B_a$ from (40) rather than $B_H$ causes $J_C/J_{SCLC}$ to have similar shapes for various $\vartheta$, indicating that much of the variation between the results for different $\vartheta$ in Fig. 6 arose because we scaled the magnetic field to $B_H$, which included geometric behavior. Because $B_H$ decreases substantially with increasing tip sharpness, examining points near $B = B_H$ requires taking small step sizes for magnetic fields in Fig. 6, which will all be close to the singularity, making calculations difficult. The planar scaling remains unchanged with $D_a = D$, permitting comparison between plots. The singularity that occurs at $B = B_H$ still occurs when scaling $B$ to $B_a$ at decreasing $B$ with decreasing $\vartheta$. Thus, while using $B_a$ for the scaling introduces nonuniversal behavior with $\vartheta$, it effectively shows more physically relevant trends.

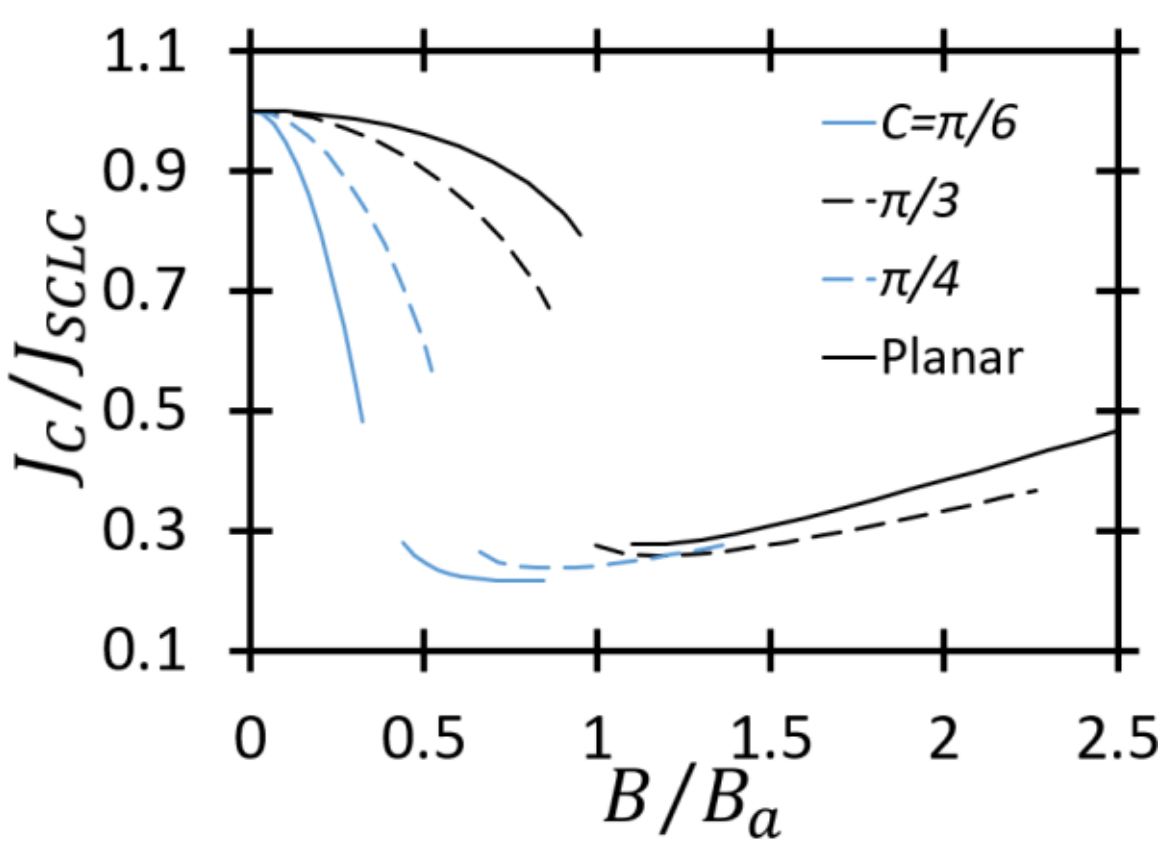

**FIG. 7.** Space-charge-limited current (SCLC) for a tip-to-tip crossed-field diode $J_C$ normalized to the SCLC for a nonmagnetic tip-to-tip diode $J_{SCLC}$ (39) as a function of magnetic field $B$ normalized to the $\vartheta$-independent $B_a$ for various $\vartheta$ with the planar result [27-30] included for comparison. Note that the curves plotted here include the same points as those from Fig. 6 where $B$ was scaled to $B_H$.

**E. Tip-to-plane geometry**

For tip-to-plane, we use the equations from tip-to-tip with $A = \pi/2$ to account for the planar anode. Figure 8 shows the three-dimensional (3D) geometry and representative cross-section and electron trajectories in a crossed-field gap.

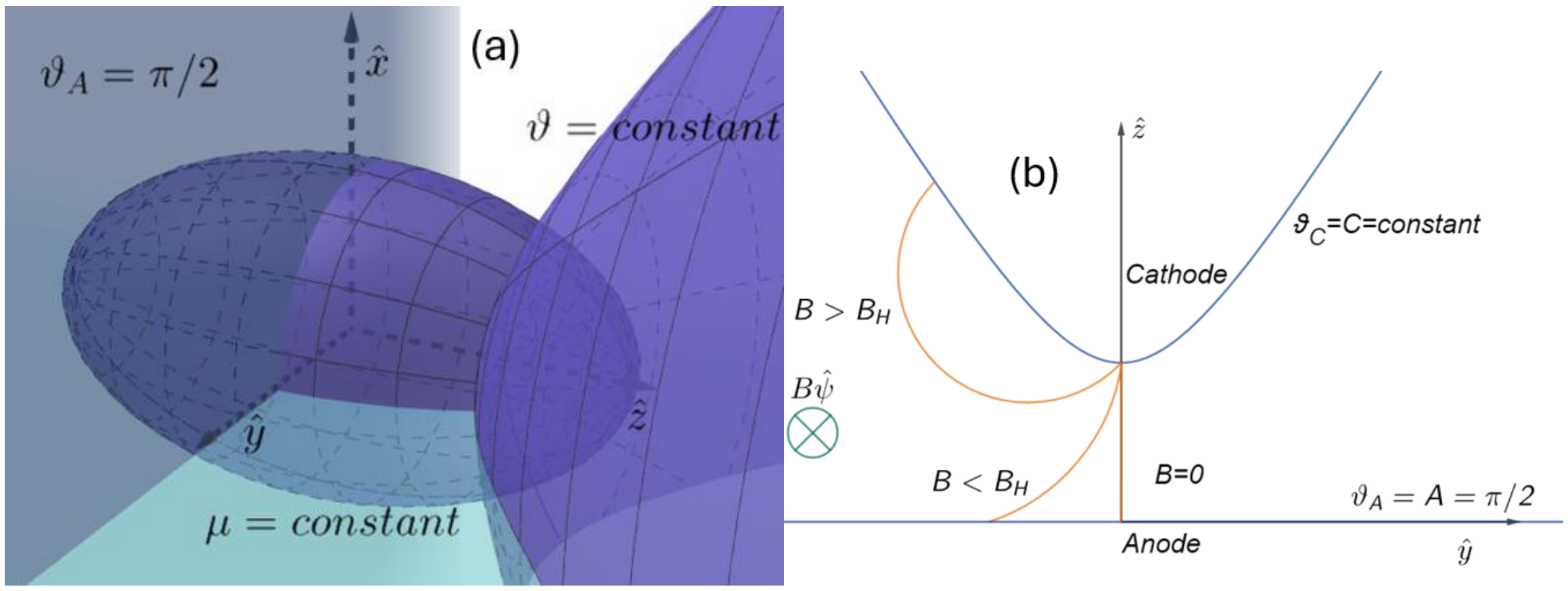

**FIG. 8.** Prolate spheroidal coordinate system used for the tip-to-plane geometry. (a) Three-dimensional (3D) geometry for a constant angle $\vartheta$ and length $\mu$ with $\nu = \pi/2$. The distance between the foci and origin $a$ is unity. The $q$ variable in this geometry is $\vartheta$. (c) Representative diode showing estimated electron trajectories for $B = 0$, $B < B_H$, and $B > B_H$.

At $A = \pi/2$, $B_H$ is given by (B4) and the electric potential by (B5). We can use the same boundary conditions as for the tip-to-tip, so $\bar{\phi}(C) = 0$ and $\bar{\phi}(A) = 1$ for $B < B_H$, $\bar{\phi}'(C) = 0$, and $\bar{\phi}'(H)$ comes from (37) evaluated with $A = \pi/2$ with the prime ($'$) representing differentiation with respect to $\vartheta$. We obtain $H$ by using (38) with $A = \pi/2$.

Figure 9 compares the SCLC for crossed-field diodes with tip-to-tip geometries scaled to the SCLC for a tip-to-tip geometry without the magnetic field, given by [9, 17, 20]

$$J_{SCLC} = \frac{4V^{3/2}\epsilon_0\sqrt{2e/m}}{9a^2\sin^4(C)\{\ln[\tan(C/2)]\}^2} \tag{41}$$

for various $\vartheta$ and $a = 1$. As in Fig. 6, for $B < B_H$ and $B \gg B_H$, $J_C/J_{SCLC}$ diverges from the planar solution with decreasing $\vartheta$. As $B \to B_H^-$, $J_C/J_{SCLC}$ converges to planar with increasing $\vartheta$.

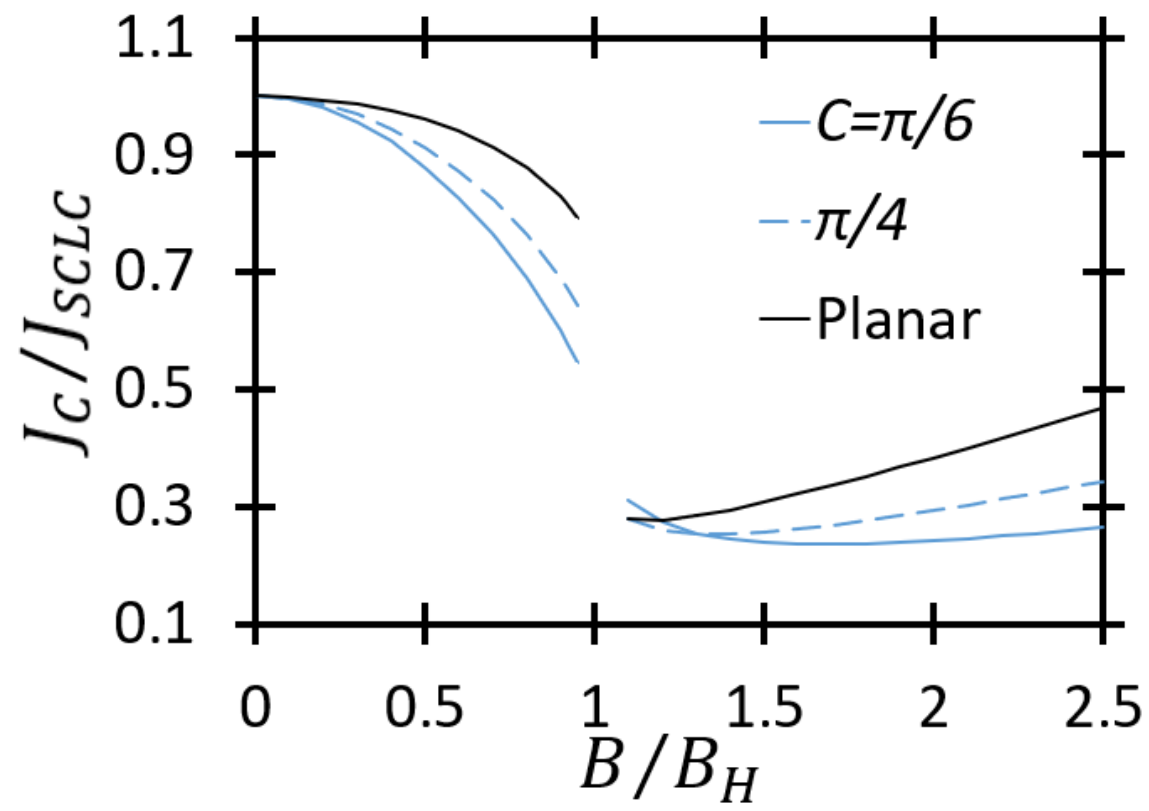

**FIG. 9.** Space-charge-limited current (SCLC) for a tip-to-plane crossed-field diode $J_C$ normalized to the SCLC for a nonmagnetic tip-to-plane diode (41) as a function of magnetic field $B$ normalized to the Hull cutoff magnetic field $B_H$ (B4) for various $\vartheta$ compared to the planar result [27-30]. Note that the curves plotted here include the same points as those from Fig. 8 where $B$ was scaled to $B_H$.

Scaling $B$ to $B_a$ from (40) rather than $B_H$ permits us to assess the impact of scaling to a $\vartheta$-independent version of the Hull cutoff in Fig. 10. Compared to the tip-to-tip geometry in Fig. 6, $B_H$ is higher for the tip-to-plane geometry (Fig. 11), eliminating the issues with the numerical solution present for the tip-to-tip geometry. Additionally, the discontinuity that arises in $J_C/J_{SCLC}$ at $B = B_H$ is less abrupt for the tip-to-plane geometry than for the planar geometry, as demonstrated by the relatively smooth decrease in $J_C/J_{SCLC}$ for $\vartheta = \pi/6$ compared to the sharp drop for the planar result at $B = B_H$. For the planar geometry, $J_C/J_{SCLC}$ increases sharply as $B \to B_H^-$ and decreases sharply as $B \to B_H^+$ [28]. While the numerical issues caused by solving near $B_H$ obfuscate these trends, the general trend from the planar case is represented in both Figs. 7 and 10 for both tip-to-tip and tip-to-plane geometries, respectively. However, the magnitude of the Hull cutoff stretches (for $B_H/B_a > 1$) or compresses ($B_H/B_a < 1$) the curve. Figure 11 shows that $B_H/B_a$ always exceeds unity for the tip-to-plane case, which makes the decrease of $J_C/J_{SCLC}$ at the discontinuity at $B = B_H$ less steep than for the planar geometry (for which $B_H = B_a$). Despite its impact on $J_C/J_{SCLC}$, the trend changes between geometries are not determined solely by $B_H$.

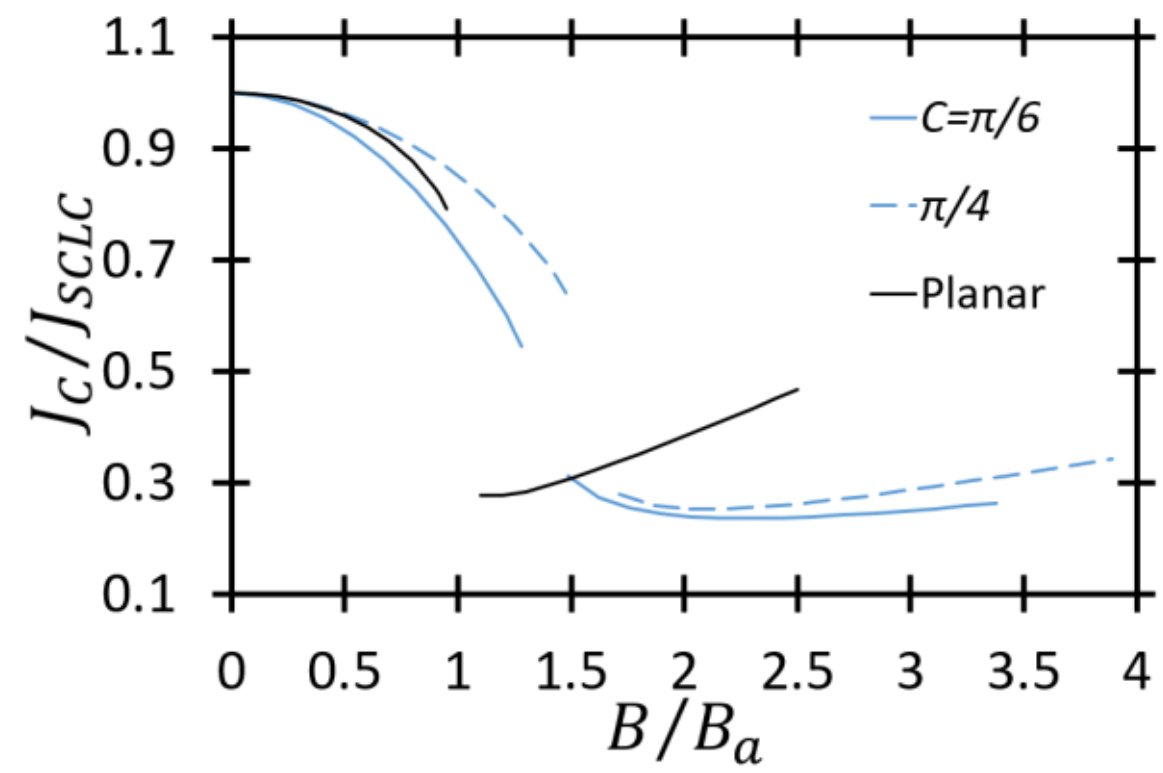

**FIG. 10.** Space-charge-limited current (SCLC) for a tip-to-plane crossed-field diode $J_C$ normalized to the SCLC for a nonmagnetic tip-to-plane diode $J_{SCLC}$ from (41) as a function of magnetic field $B$ normalized to the $\vartheta$-independent $B_a$ for various $\vartheta$-with the planar result [27-30] included for comparison.

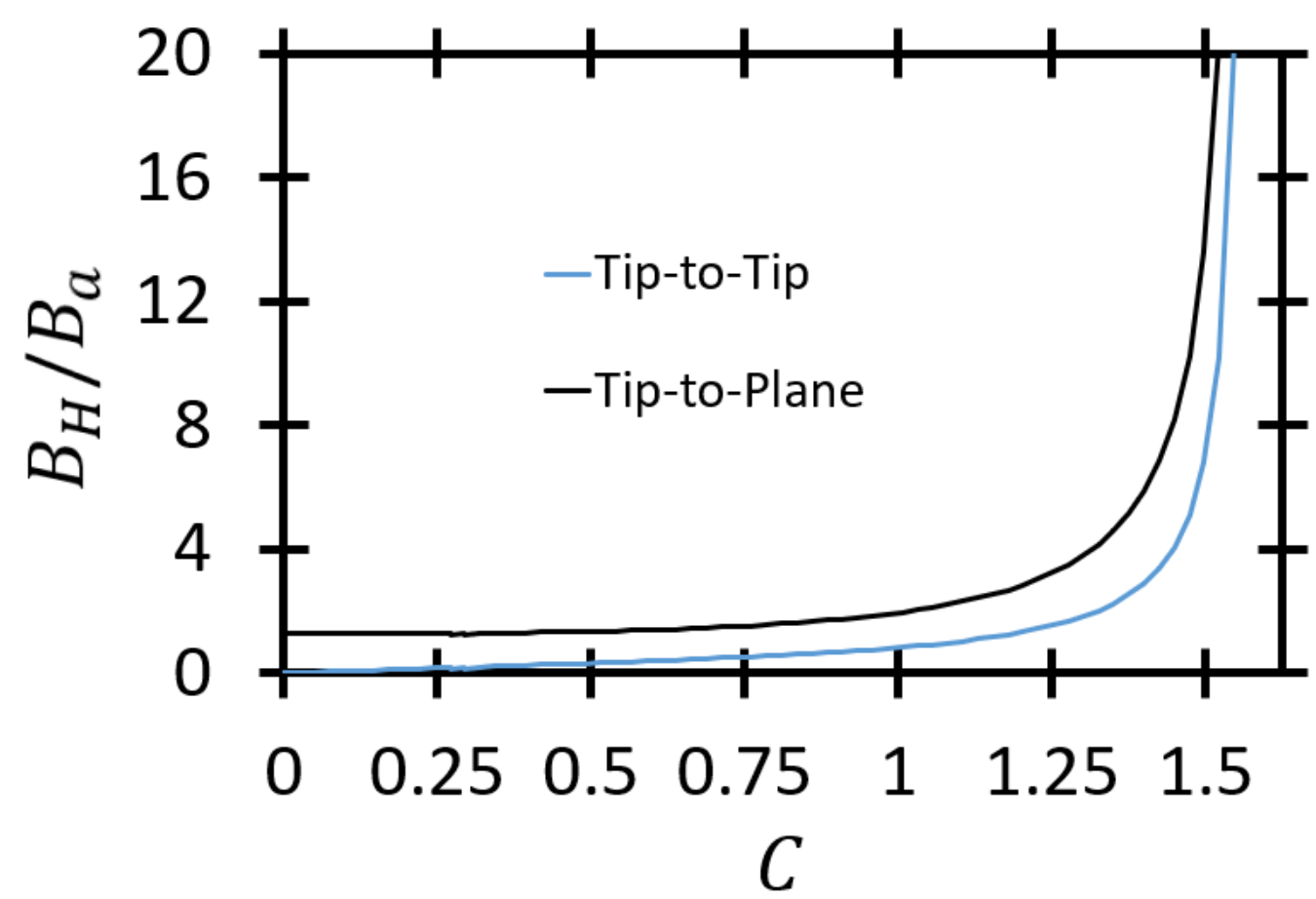

**FIG. 11.** The Hull cutoff magnetic field $B_H$ normalized to the $\vartheta$-independent $B_a$ as a function of the cathode angle $C$ from 0 to $\pi/2$ for tip-to-tip and tip-to-plane geometries.

**F. Generalizing geometry**

We defined $B_a$ to remove the geometric effects of $B_H$ normalization. The critical current for tip-to-tip and tip-to-plane geometries for a given $B_H/B_a$ disagree. Similar differences also arise for $J_C/J_{SCLC}$ in spherical and cylindrical geometries from Fig. 4 and Ref. [47], respectively, even though they have identical $B_H$. This indicates that $B_H$ and $B_a$ do not fully generalize $J_{SCLC}$ for geometry, so a coordinate-system invariant solution requires another approach. The most common approaches for assessing SCLC for nonplanar geometries in the literature are conformal mapping [16] and point transformations [17, 18, 20], which ultimately provide a canonical gap distance $\mathcal{D}$ for determining the SCLC using electric potential. The SCLC *without* a magnetic field is the same for any $\mathcal{D}$ regardless of geometry with Table I summarizing $\mathcal{D}$ for the geometries considered here. Using this same concept, we define a "canonical magnetic field" as

$$B_{\mathcal{D}} = \sqrt{\frac{2mV}{e\mathcal{D}^2}} = \sqrt{\frac{2mV}{e}}\frac{1}{|\mathcal{D}|}. \tag{42}$$

Equation (42) differs from (18), which we derived using scale factors, which gave $D_M$ in the denominator. Table II shows $\mathcal{D}$ and $\delta = \mathcal{D}/D_M = B_H/B_{\mathcal{D}}$ for the geometries considered here. Note that $\mathcal{D}$ accounts for the geometric effects resulting from electric potential, while $D_M$ resolves geometric effects resulting from the crossed-field.

TABLE II: Canonical gap distance $\mathcal{D}$ and the scaling between the Hull cutoff magnetic field $B_H$ using the appropriate scale factors and the "canonical magnetic field" $B_{\mathcal{D}}$ using the appropriate canonical gap distance.

| Gap type | Canonical Gap Distance ($\mathcal{D}$) | $\delta = B_H/B_{\mathcal{D}} = \mathcal{D}/D_M$ |
|---|---|---|
| Flat Plate | $A - C$ | 1 |
| Concentric Cylinders | $R_C \ln(R_A/R_c)$ | $\frac{2\lvert\ln(a_r)\rvert}{a_r\lvert a_r^{-2} - 1\rvert}$ |
| Concentric Spheres | $R_C(R_A - R_C)/R_A$ | $\frac{2\lvert(1 - a_r)\rvert}{a_r\lvert(a_r^{-2} - 1)\rvert}$ |
| Rounded Tip-to-Tip | $a\sin^2(C)\ln\left[\frac{\tan(A/2)}{\tan(C/2)}\right]$ | $\left\lvert\frac{2\sin(A)\sin^2(C)\ln\left[\frac{\tan(A/2)}{\tan(C/2)}\right]}{(A - C - \cos(A)\sin(A) + \cos(C)\sin(C))}\right\rvert$ |

Figure 12 shows $J_C/J_{SCLC}$ as a function of $B/B_{\mathcal{D}}$ from (48) instead of $B/B_H$ from (18). We note clear trends in $J_C/J_{SCLC}$ with geometry in Fig. 12 that were obfuscated by the scaling when we plotted as a function of $B/B_H$. Universally, the singularity created by the turning of the electrons in terms of $B/B_{\mathcal{D}}$ shifts left as the geometry becomes less planar. For $B/B_{\mathcal{D}}$ beyond the singularity, $J_C/J_{SCLC}$ increases more steeply for cylindrical and spherical geometries than for either tip geometry. Moreover, for $B/B_{\mathcal{D}} \gg 1$, $J_C/J_{SCLC}$ becomes independent of tip angle for both tip-to-tip and tip-to-plane

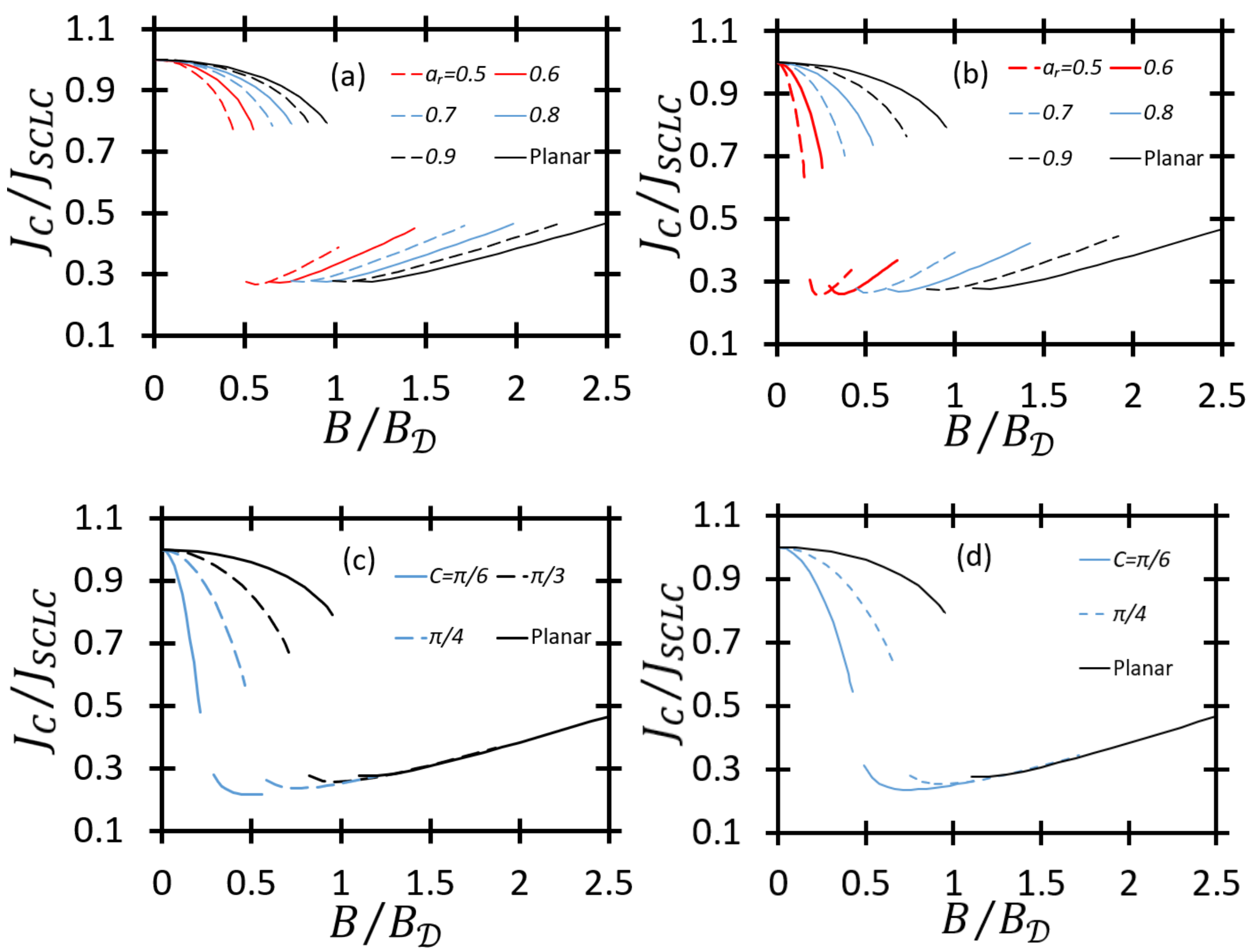

**FIG. 12.** Space-charge-limited current (SCLC) for a crossed-field diode $J_C$ scaled to the SCLC for a nonmagnetic diode $J_{SCLC}$ as a function of the magnetic field $B$ scaled to the geometry generalized Hull cutoff magnetic field $B_{\mathcal{D}}$ for (a) cylindrical, (b) spherical, (c) tip-to-tip, and (d) tip-to-plane geometries for various $a_r = R_C/R_A$ for cylinders and spheres, where the cathode radius is $R_C$ and the anode radius is $R_C$, and cathode angle $C$ for tip-to-tip and tip-to-plane.

We next attempt to derive a scaling factor that would yield a universal curve of $J_C/J_{SCLC}$ across all geometries. To derive such a universal relationship, we define $\delta = B_H/B_{\mathcal{D}} = \mathcal{D}/D_M$ for the geometries from Table II for $0 \leq C \leq \pi/2$ (or $0 \leq a_r \leq 1$). While singularities occur at $C = 0$ and $C = \pi/2$ for the tip geometries and $a_r = 1$ and $a_r = 0$ for the cylindrical and spherical geometries, we can determine these limits by employing series expansions or L'Hôpital's rule. As $a_r \to 1$ or $C \to \pi/2$, $\bar{J}_C/\bar{J}_{SCLC}$ for the cylindrical [47], spherical (Fig. 4), tip-to-tip (Fig. 6), and tip-to-plane (Fig. 9) geometries approach the planar solution. Since the boundary condition does not prove this in general, we show $\delta$ as a function of $C$ for the tip geometries (or $a_r$ for the cylindrical and spherical geometries) in Fig. 13 to demonstrate that $\delta \to 1$ as $C \to \pi/2$ or $a_r \to 1$. We next test our conjecture that $\bar{J}_C/\bar{J}_{SCLC}$ should match across geometries for a given $\delta$, particularly as $\delta \to 1$.

Figure 14 shows that $\bar{J}_C/\bar{J}_{SCLC}$ remains remarkably unchanged (although *not* identical) across geometry when we consider a specific $\delta = \mathcal{D}/D_M$ for $\delta \geq 0.6$. As $\delta \to 1$, the results for the different geometries overlap since they are geometrically approaching the planar solution, which always has $\delta = 1$ (Table II). We can also consider this as aligning the singularities from Fig. 12 by accounting for the geometric effects of the electric field through $\mathcal{D}$ and the geometric effects from the magnetic field through $D_M$ to determine "equivalent" dimensions in the various geometries. The imperfect agreement between the various geometries at lower $\delta$ indicates that an additional geometric correction, which is most likely a nonlinear function of $\mathcal{D}$ and/or $D_M$, must contribute. This agreement between all geometries for $\delta \geq 0.6$ in this limiting case provides additional confidence in the relative accuracy of these results.

This difference is analogous to our prior derivation of the uniform SCLC for a two-dimensional planar diode with nonzero monoenergetic initial velocity [49], where we demonstrated analytically that the contributions of nonzero initial velocity and geometry could not be decoupled for higher levels of initial velocity. In that case, we observed that we could apply a simple semi-empirical correction that considered the corrections of geometry and initial velocity as multiplicative correction factors with excellent agreement at lower initial velocities and diminishing agreement with increasing initial velocity [49]. We speculate that crossed-field behavior is similarly coupled and completely incorporating the individual contributions of electric and magnetic fields would require revisiting the full derivation. Moreover, we have made the analogy here to our prior studies of canonical gap distance $\mathcal{D}$; however, $\delta = 1$ for planar diodes due our definitions of the scaling parameters, while $0 \leq \delta \leq 1$ for the other geometries. This prevents complete generalization of any other case to the planar case as the planar $\delta$ is fixed, although the other geometries approach planar as $\delta \to 1$. Again, revisiting the initial steps of the derivation may elucidate this behavior; however, for now, we have taken a first step toward illustrating the intricacies of the contributions of geometry on magnetic insulation and critical current in a crossed-field diode.

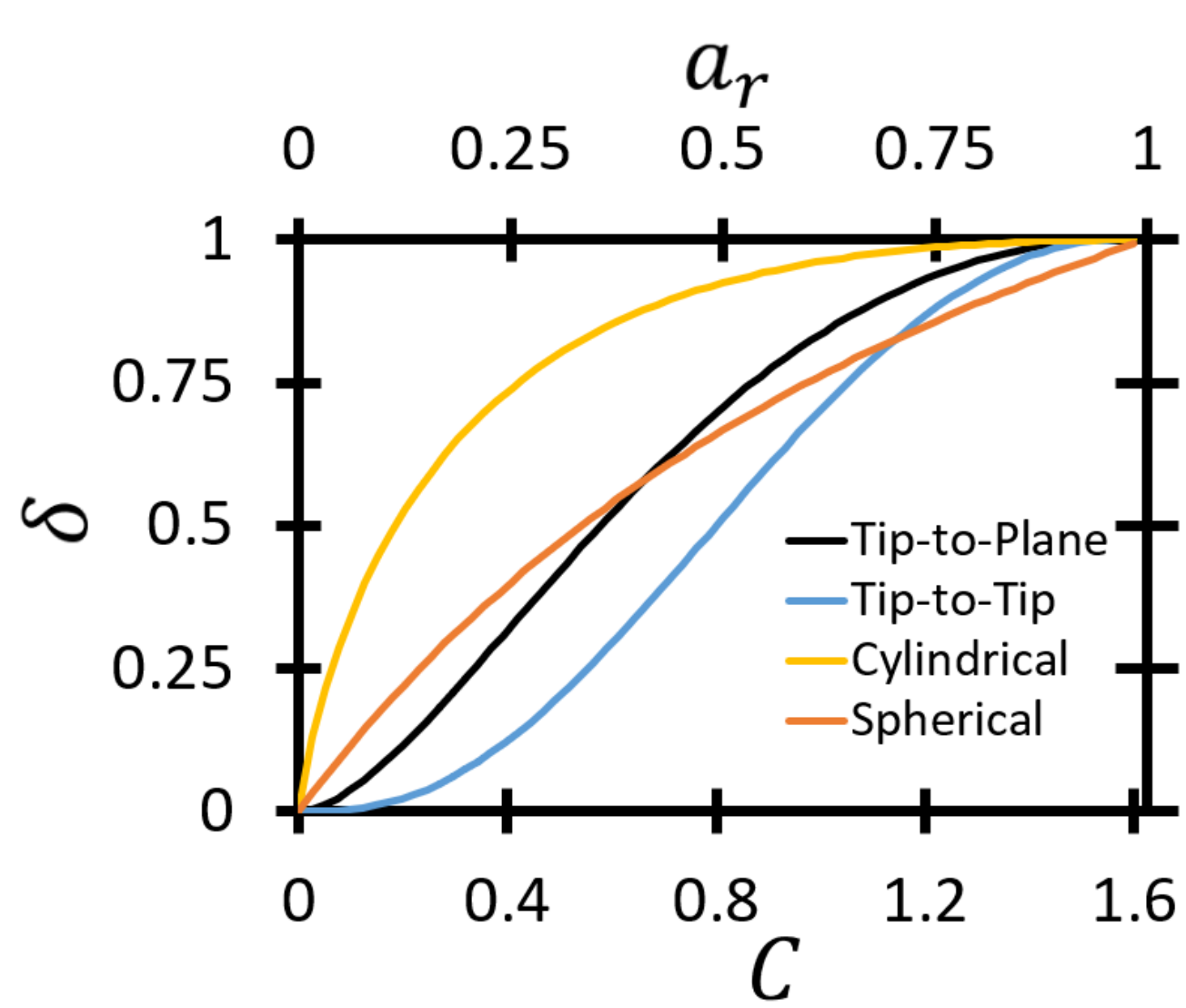

**FIG. 13.** Assessment of $\delta = \mathcal{D}/D_M$, which describes the scaling of the contribution of geometric effects of the electric field given by the canonical gap distance $\mathcal{D}$ to the geometric effects of the crossed-field defined by $D_M$, as a function of cathode angle $C$ for the tip-to-tip and tip-to-plane geometries or $a_r$ for the cylindrical and spherical geometries. As $C \to \pi/2$ or $a_r \to 1$, $\delta \to 1$, indicating $\mathcal{D} \approx D_M$, indicating that the geometries are approaching planar. This curve is mirrored from $\pi/2 < C < \pi$ (or $1 < a_r < \infty$), meaning that $\delta \to 0$ as $C \to \pi$ (or $a_r = R_C/R_A \to \infty$).

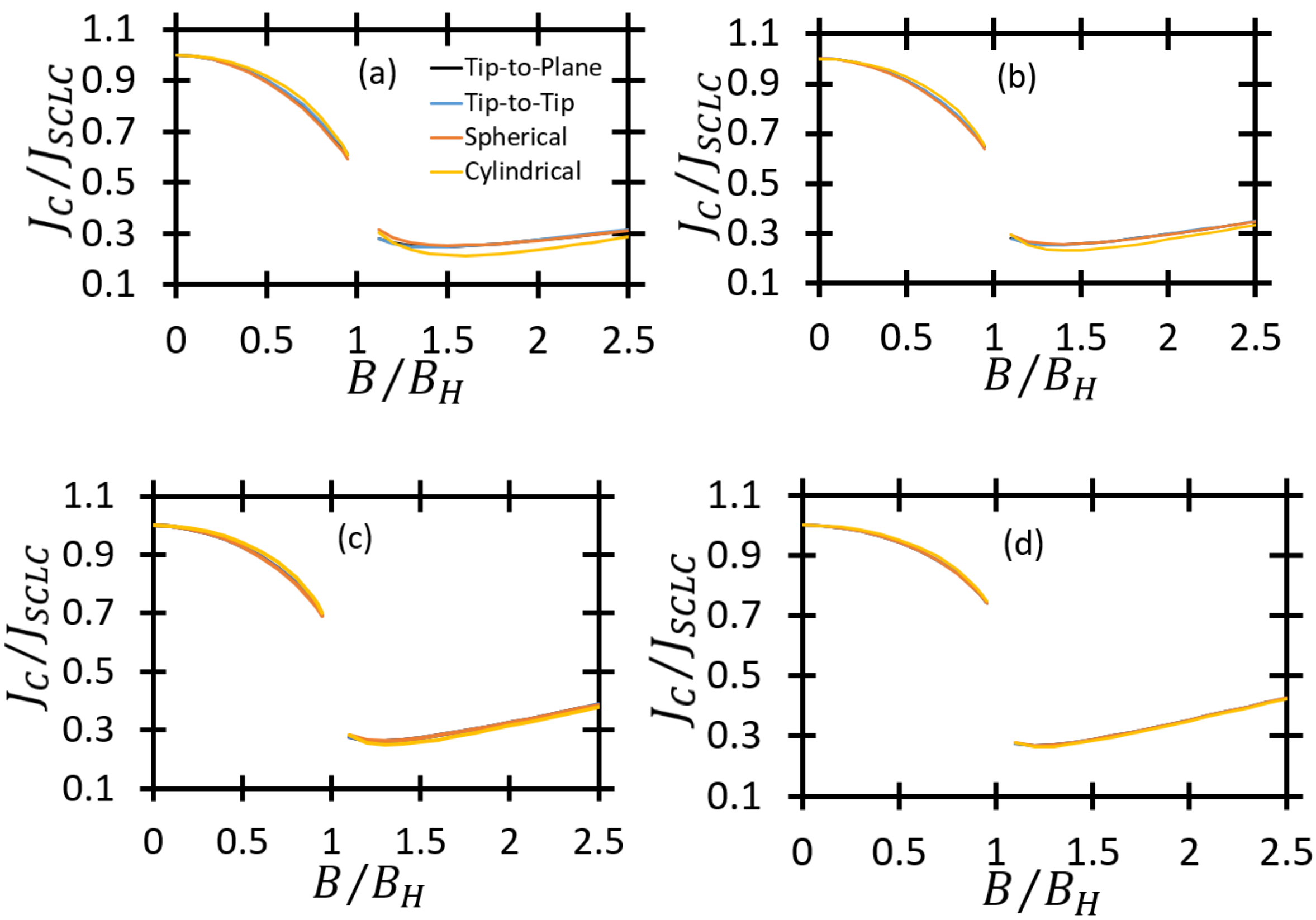

**FIG. 14.** Assessment of universal scalings of $J_C/J_{SCLC}$ as a function of $B/B_H$ for $\delta = B_H/B_{\mathcal{D}} = \mathcal{D}/D_M$ of (a) 0.6, (b) 0.7, (c) 0.8, and (d) 0.9 for each geometry in Fig. 9, corresponding to(a) $a_{r,cyl} = 0.159, a_{r,sph} = 0.429, C_{ttt} = 0.896, C_{ttp} = 0.685,$ (b) $a_{r,cyl} = 0.220, a_{r,sph} = 0.538, C_{ttt} = 0.997, C_{ttp} = 0.803,$ (c) $a_{r,cyl} = 0.306, a_{r,sph} = 0.667, C_{ttt} = 1.11, C_{ttp} = 0.943$ and (d) $a_{r,cyl} = 0.448, a_{r,sph} = 0.818, C_{ttt} = 1.25, C_{ttp} = 1.23$. The agreement of all four curves indicates the relative coordinate system invariance over this range of $\delta$, particularly as $\delta \to 1$, which corresponds to the geometries approaching planar.

## IV. CONCLUSION

We have derived the SCLC for nonrelativistic, vacuum crossed-field diodes with electrons emitted with zero initial velocity that can be solved numerically for *any* orthogonal coordinate system by applying the result from numerically evaluating (19) to (16). We compared the results to prior solutions obtained using variational calculus for planar and cylindrical diodes. We then extended this solution to spherical, tip-to-tip, and tip-to-plane geometries for both $B < B_H$ and $B > B_H$. This gives the maximum current density permissible for *any* crossed-field diode with an orthogonal coordinate system.

In the process, we derived various magnetic field (and, concomitantly, length) scalings to physically describe our results, as summarized mathematically in Table I and physically in Table III. We derived an equation for $B_H$ (18) that holds for any orthogonal geometry and showed that $B_H \to 0$ for a tip-to-tip geometry as the tips become infinitely sharp. The rapidly decreasing $B_H$ made plotting $J_C/J_{SCLC}$ as a function of $B/B_H$ numerically challenging, motivating us to change the scaling to a geometrically independent magnetic field. Interestingly, we did not encounter the same issue with the tip-to-plane geometry since $B_H$ was larger.

TABLE III: Comparison of different scaling factors for the magnetic field used throughout this study.

| Magnetic Field Scaling | Description | Purpose |
|---|---|---|
| $B_H$ | Magnetic field required for the electron to reach the anode with zero velocity in the direction of the electric field. | Evaluate the SCLC with magnetic field scaled to the condition for magnetic insulation. |
| $B_a$ | Magnetic field scaling independent of the angle factors in tip geometries | Evaluate the trends as $B \to B_H$ to determine the trends arising due to approaching magnetic insulation and those due to angles introduced by the tip(s). |
| $B_{\mathcal{D}}$ | Magnetic field scaled to the canonical gap distance $\mathcal{D}$ based on mapping the electric potential. | Examine the feasibility of combining the magnetic field scaling from $B_H$ and the electric field |

scaling from
$B_{\mathcal{D}}$ to account for both phenomena.

Scaling the magnetic field to $B_a$ revealed that the poor convergence as $B \to B_H$ for the tip geometries arose because the low $B_H$ required taking increasingly small steps that were not computationally achievable. This motivated the search for a geometrically universal solution for $J_C/J_{SCLC}$ by defining a universal parameter $\delta = \mathcal{D}/D_M$, where $\mathcal{D}$ is the canonical gap distance from previous studies of SCLC without a magnetic field and $D_M$ accounts for the geometry for the crossed-field. The SCLC for the four geometries considered agreed well when we selected their dimensions to achieve a constant $\delta \geq 0.6$, overlapping as $\delta \to 1$, which corresponds to a planar geometry. While the current approach generally requires a higher (more planar) $\delta$, it suggests the feasibility of developing a truly coordinate system invariant solution for SCLC for a crossed-field diode that more completely couples geometry and magnetic field effects. Since variational calculus, conformal mapping, and point transformations are all based on examining potential, we conjecture that considering the magnetic vector potential, which is infrequently used in crossed-field studies [43], may facilitate the connection of these phenomena. This could allow us to extend these calculations to geometries commonly used for operational crossed-field devices, such as slow wave structures [50], if we can further incorporate multidimensional effects [51] as we have done without the crossed-field [22].

**ACKNOWLEDGMENTS**

This research was supported by the Air Force Office of Scientific Research under Grant Number FA9550-22-1-0434. We also gratefully acknowledge Adam Darr for useful discussions.

## AUTHOR DECLARATIONS

### Conflict of Interest

The authors have no conflicts to disclose.

### Author Contributions

**Jack K. Wright:** Conceptualization (supporting), Formal Analysis (lead), Investigation (lead), Methodology (lead), Writing – original draft (lead), Writing – review and editing (supporting); **N. R. Sree Harsha:** Conceptualization (supporting), Formal Analysis (supporting), Investigation (supporting), Methodology (supporting), Writing – review and editing (supporting). **Allen L. Garner:** Conceptualization (lead); Formal Analysis (supporting); Funding acquisition (lead); Investigation (supporting); Methodology (supporting); Project Administration (lead); Supervision (lead); Writing – original draft (supporting); Writing – review & editing (lead).

## DATA AVAILABILITY

The data that support the findings of this study are available from the corresponding author upon reasonable request.

## APPENDIX A: DERIVATION OF THE GENERAL SOLUTION

Evaluating the integrand $f$ from (17) in (15) provides an additional equation completely describing the electric potential for $B < B_H$ and the electric potential in the region containing space-charge for $B > B_H$ as

$$
\begin{aligned}
&\frac{\partial^2}{\partial q^2}\frac{\partial f}{\partial \phi''} - \frac{\partial}{\partial q}\frac{\partial f}{\partial \phi'} + \frac{\partial f}{\partial \phi} \\
&= \frac{-2B^2 e m h_1^3 h_2^6 h_3 \phi + h_2 h_3 \left(-2m h_2^2 \phi + B^2 e\left(\int_C^q h_1 h_2 dq\right)^2\right) h_1' \left(B^2 e\left(\int_C^q h_1 h_2 dq\right)^2 h_2' + 2m h_2^3 \phi'\right)}{h_1 h_2 \left(2m h_2^2 \phi - B^2 e\left(\int_C^q h_1 h_2 dq\right)^2\right)^{3/2}} \\
&+ \frac{B^2 e h_1^2 h_2^3 \int_C^q h_1 h_2 dq \left(4m h_2 h_3 \phi h_2' + B^2 e\left(\int_C^q h_1 h_2 dq\right)^2 h_3' + h_2^2\left(-2m\phi h_3' + 2m h_3 \phi'\right)\right)}{h_1 h_2 \left(2m h_2^2 \phi - B^2 e\left(\int_C^q h_1 h_2 dq\right)^2\right)^{3/2}} \\
&- \frac{4B^2 e m h_2^2 h_3 \phi\left(\int_C^q h_1 h_2 dq\right)^2 h_2'^2 - B^4 e^2 h_3\left(\int_C^q h_1 h_2 dq\right)^4 h_2'^2 - 4m^2 h_2^5 h_3 \phi h_2' \phi'}{h_2 \left(2m h_2^2 \phi - B^2 e\left(\int_C^q h_1 h_2 dq\right)^2\right)^{3/2}} \\
&- \frac{2B^2 e m h_2^3\left(\int_C^q h_1 h_2 dq\right)^2 \left(-2h_3 h_2' \phi' + \phi\left(h_2' h_3' + h_3 h_2''\right)\right) + 2B^2 e m h_2^3\left(\int_C^q h_1 h_2 dq\right)^2 \left(h_3' \phi' + h_3 \phi''\right)}{h_2 \left(2m h_2^2 \phi - B^2 e\left(\int_C^q h_1 h_2 dq\right)^2\right)^{3/2}} \\
&+ \frac{m^2 h_2^6 \left(h_3 \phi'^2 - 4\phi\left(h_3' \phi' + h_3 \phi''\right)\right) + B^4 e^2 h_2\left(\int_C^q h_1 h_2 dq\right)^4 \left(h_2' h_3' + h_3 h_2''\right)}{h_2 \left(2m h_2^2 \phi - B^2 e\left(\int_C^q h_1 h_2 dq\right)^2\right)^{3/2}} = 0.
\end{aligned}
\tag{A1}
$$

Solving (A1) for $\phi''$ yields

$$
\begin{aligned}
&\phi'' \\
&= \frac{2B^2 e m h_1^3 h_2^6 h_3 \phi + h_2 h_3 \left(2m h_2^2 \phi - B^2 e\left(\int_C^q h_1 h_2 dq\right)^2\right) h_1' \left(B^2 e\left(\int_C^q h_1 h_2 dq\right)^2 h_2' + 2m h_2^3 \phi'\right)}{2m h_1 h_2^4 h_3 \left(2m h_2^2 \phi - B^2 e\left(\int_C^q h_1 h_2 dq\right)^2\right)} \\
&- \frac{B^2 e h_1^2 h_2^3 \int_C^q h_1 h_2 dq \left(4m h_2 h_3 \phi h_2' + B^2 e\left(\int_C^q h_1 h_2 dq\right)^2 h_3' + h_2^2\left(-2m\phi h_3' + 2m h_3 \phi'\right)\right)}{2m h_1 h_2^4 h_3 \left(2m h_2^2 \phi - B^2 e\left(\int_C^q h_1 h_2 dq\right)^2\right)} \\
&+ \frac{4B^2 e m h_2^2 h_3 \phi\left(\int_C^q h_1 h_2 dq\right)^2 h_2'^2 - B^4 e^2 h_3\left(\int_C^q h_1 h_2 dq\right)^4 h_2'^2 - 4m^2 h_2^5 h_3 \phi h_2' \phi' + 2B^2 e m h_2^4\left(\int_C^q h_1 h_2 dq\right)^2 h_3' \phi'}{2m h_2^4 h_3 \left(2m h_2^2 \phi - B^2 e\left(\int_C^q h_1 h_2 dq\right)^2\right)} \\
&+ \frac{m^2 h_2^6 \phi'\left(-4\phi h_3' + h_3 \phi'\right) + B^4 e^2 h_2\left(\int_C^q h_1 h_2 dq\right)^4 \left(h_2' h_3' + h_3 h_2''\right)}{2m h_2^4 h_3 \left(2m h_2^2 \phi - B^2 e\left(\int_C^q h_1 h_2 dq\right)^2\right)} \\
&- \frac{2B^2 e m h_2^3\left(\int_C^q h_1 h_2 dq\right)^2 \left(\phi h_2' h_3' - 2h_3 h_2' \phi' + h_3 \phi h_2''\right)}{2m h_2^4 h_3 \left(2m h_2^2 \phi - B^2 e\left(\int_C^q h_1 h_2 dq\right)^2\right)}.
\end{aligned}
\tag{A2}
$$

Substituting (A2) into the general equation for the Laplacian (1) and simplifying gives the general form that holds for any orthogonal coordinate system [cf., (19) in the main text].

## APPENDIX B: DERIVATION OF TIP-TO-TIP AND TIP-TO-PLANE CROSSED-FIELD SOLUTION

For the tip-to-tip and tip-to-plane geometries, the 1D scale factors, $h_1 = h_2 = h_3 = a\sin(\vartheta)$, where $a$ is the distance between the foci and the origin, into the general form (19) to obtain

$$\nabla^2\phi = \csc^6(\vartheta)\left[\frac{32a^2B^2em\phi\sin^8(\vartheta) - a^4B^4e^2\cos^2(\vartheta)\,(C-\vartheta-\cos(C)\sin(C)+\cos(\vartheta)\sin(\vartheta))^4}{32m\left(2a^2m\phi\sin^2(\vartheta) - \frac{1}{4}a^4B^2e(C-\vartheta-\cos(C)\sin(C)+\cos(\vartheta)\sin(\vartheta))^2\right)}\right.$$
$$+\frac{(C-\vartheta-\cos(C)\sin(C)+\cos(\vartheta)\sin(\vartheta))^2\left(2a^2m\phi\sin^2(\vartheta) - \frac{1}{4}a^4B^2e(C-\vartheta-\cos(C)\sin(C)+\cos(\vartheta)\sin(\vartheta))^2\right)}{\frac{8m}{eB^2\cos^2(\vartheta)}\left(2a^2m\phi\sin^2(\vartheta) - \frac{1}{4}a^4B^2e(C-\vartheta-\cos(C)\sin(C)+\cos(\vartheta)\sin(\vartheta))^2\right)}$$
$$+\frac{a^2B^2em\phi\sin^2(2\vartheta)\,(2C-2\vartheta-\sin(2C)+\sin(2\vartheta))^2 + \frac{1}{16}a^4B^4e^2\cos(2\vartheta)\,(2C-2\vartheta-\sin(2C)+\sin(2\vartheta))^4}{32m\left(2a^2m\phi\sin^2(\vartheta) - \frac{1}{4}a^4B^2e(C-\vartheta-\cos(C)\sin(C)+\cos(\vartheta)\sin(\vartheta))^2\right)}$$
$$+\frac{a^2B^2e\sin^2(nu)\,(2C-2\vartheta-\sin(2C)+\sin(2\vartheta))^2(\sin(2\vartheta)\,\phi' - 2\cos(2\vartheta)\,\phi)}{32m\left(2a^2m\phi\sin^2(\vartheta) - \frac{1}{4}a^4B^2e(C-\vartheta-\cos(C)\sin(C)+\cos(\vartheta)\sin(\vartheta))^2\right)}$$
$$+\frac{8m^2\sin^6(\vartheta)\,\phi'^2}{16m\left(2a^2m\phi\sin^2(\vartheta) - \frac{1}{4}a^4B^2e(C-\vartheta-\cos(C)\sin(C)+\cos(\vartheta)\sin(\vartheta))^2\right)}$$
$$-\frac{a^2B^2e\sin^3(\vartheta)\,(\vartheta - C+\cos(C)\sin(C)-\cos(\vartheta)\sin(\vartheta))}{16m\left(2a^2m\phi\sin^2(\vartheta) - \frac{1}{4}a^4B^2e(C-\vartheta-\cos(C)\sin(C)+\cos(\vartheta)\sin(\vartheta))^2\right)}$$
$$\times\left[(a^2B^2e\cos(\vartheta)\,(C-\vartheta-\cos(C)\sin(C)+\cos(\vartheta)\sin(\vartheta))^2 + 8m\phi\sin(\vartheta)\sin(2\vartheta)\right.$$
$$\left.\left. + 8m\sin^2(\vartheta)\,(\sin(\vartheta)\,\phi' - \cos(\vartheta)\,\phi)\right]\right]. \tag{B1}$$

where the prime ($'$) denotes differentiation with respect to $\vartheta$. Using $\bar{\phi} = \phi/V$ and defining the Hull cutoff magnetic field for a tip-to-tip geometry $B_{H,ttt}$ from (18) as

$$B_{H,ttt} = \sqrt{\frac{2mV}{e}}\left|\frac{2\sin(A)}{a(A-C-\cos(A)\sin(A)+\cos(C)\sin(C))}\right| \tag{B2}$$

allows us to recast (B1) as

$$
\begin{aligned}
&\nabla^2\bar{\phi} \\
&= \frac{\csc^2(\vartheta)}{32a^2(A-C-\cos(A)\sin(A)+\cos(C)\sin(C))^2} \\
&\times\Bigg[\frac{-\bar{B}^4\csc^6(\vartheta)\sin^4(A)\,(2C-2\vartheta-\sin(2C)+\sin(2\vartheta))^3(4C-4\vartheta-2\sin(2C)+\sin(4\vartheta))}{(\bar{\phi}(2A-2C-\sin(2A)+\sin(2C))^2-\bar{B}^2\csc^2(\vartheta)\sin^2(A)\,(2C-2\vartheta-\sin(2C)+\sin(2\vartheta))^2)} \\
&-\frac{\bar{B}^2\csc^4(\vartheta)\,\bar{\phi}\sin^2(\vartheta)\,(2A-2C-\sin(2A)+\sin(2C))^2}{(\bar{\phi}(2A-2C-\sin(2A)+\sin(2C))^2-\bar{B}^2\csc^2(\vartheta)\sin^2(A)\,(2C-2\vartheta-\sin(2C)+\sin(2\vartheta))^2)} \\
&\times\big[96C\vartheta-56-48C^2+6\cos(4C)+\cos(4C-2\vartheta)+16\cos\big(2(C-\vartheta)\big)+59\cos(2\vartheta)-16C^2\cos(2\vartheta) \\
&+32C\vartheta\cos(2\vartheta)-16\vartheta^2\cos(2\vartheta)-14\cos(4\vartheta)+3cos(6\vartheta)-16\cos\big(2(C+\vartheta)\big)+\cos\big(2(2C+\vartheta)\big) \\
&+48C\sin(2C)-48\vartheta\sin(2C)+8C\sin\big(2(C-\vartheta)\big)-8\vartheta\sin\big(2(C-\vartheta)\big)-64C\sin(2\vartheta)+64\vartheta\sin(2\vartheta) \\
&+8C\sin\big(2(C+\vartheta)\big)-8\vartheta\sin\big(2(C+\vartheta)\big)\big] \\
&+\frac{2\bar{B}^2\csc^3(\vartheta)\sin^2(A)\,(2A-2C-\sin(2A)+\sin(2C))^2\bar{\phi}'}{(\bar{\phi}(2A-2C-\sin(2A)+\sin(2C))^2-\bar{B}^2\csc^2(\vartheta)\sin^2(A)\,(2C-2\vartheta-\sin(2C)+\sin(2\vartheta))^2)} \\
&\times[8\cos(\vartheta)-8\cos(2C-\vartheta)-\cos(4C-\vartheta)+16C^2\cos(\vartheta)-32C\vartheta\cos(\vartheta)+16\vartheta^2\cos(\vartheta)-7\cos(3\vartheta) \\
&+\cos(5\vartheta)+8\cos(2C+\vartheta)-\cos(4C+\vartheta)-8C\sin(2C-\vartheta)+8\vartheta\sin(2C-\vartheta)+32C\sin(\vartheta)-32\vartheta\sin(\vartheta) \\
&-8C\sin(2C+\vartheta)+8\vartheta\sin(2C+\vartheta)] \\
&+\frac{2(2A-2C-\sin(2A)+\sin(2C))^4\bar{\phi}'^2}{(\bar{\phi}(2A-2C-\sin(2A)+\sin(2C))^2-\bar{B}^2\csc^2(\vartheta)\sin^2(A)\,(2C-2\vartheta-\sin(2C)+\sin(2\vartheta))^2)}\Bigg],
\end{aligned}
\tag{B3}
$$

To evaluate $\nabla^2\phi$ for a tip-to-plane, geometry, we substitute $A=\pi/2$ into (B2) and (B3) to obtain $B_H$ for the tip-to-plane geometry as

$$
B_{H,ttp}=\sqrt{\frac{2mV}{e}}\left|\frac{4}{a[\pi-2C+\sin(2C)]}\right|. \tag{B4}
$$

and the electric potential as

$$
\begin{aligned}
&\nabla^2\bar{\phi}\\
&=\frac{-2\bar{B}^4\csc^8(\vartheta)(2C-2\vartheta-\sin(2C)+\sin(2\vartheta))^3\,(4C-4\vartheta-2\sin(2C)+\sin(4\vartheta))}{4(\pi-2C+\sin(2C))^2(\bar{\phi}(\pi-2C+\sin(2C))^2-\bar{B}^2\csc^2(\vartheta)\,(2C-2\vartheta-\sin(2C)+\sin(2\vartheta))^2)}\\
&-\frac{\frac{1}{2}\bar{B}^2\csc^6(\vartheta)\,\bar{\phi}(\pi-2C+\sin(2C))^2}{4(\pi-2C+\sin(2C))^2(\bar{\phi}(\pi-2C+\sin(2C))^2-\bar{B}^2\csc^2(\vartheta)\,(2C-2\vartheta-\sin(2C)+\sin(2\vartheta))^2)}\\
&\times\big[96C\vartheta-56-48C^2-48\vartheta^2+6\cos(4C)+\cos(4C-2\vartheta)+16\cos\big(2(C-\vartheta)\big)+59\cos(2\vartheta)-16C^2\cos(2\vartheta)\\
&+32C\vartheta\cos(2\vartheta)-16\vartheta^2\cos(2\vartheta)-14\cos(4\vartheta)+3\cos(6\vartheta)-16\cos\big(2(C+\vartheta)\big)+\cos\big(2(2C+\vartheta)\big)\\
&+48C\sin(2C)-48\vartheta\sin(2C)+8C\sin\big(2(C-\vartheta)\big)-8\vartheta\sin\big(2(C-\vartheta)\big)-64C\sin(2\vartheta)+64\vartheta\sin(2\vartheta)\\
&+8Csin\big(2(C+\vartheta)\big)-8\vartheta\sin\big(2(C+\vartheta)\big)\big]\\
&+\frac{\bar{B}^2\csc^5(\vartheta)(\pi-2C+\sin^2(2C)\,\bar{\phi}'}{4(\pi-2C+\sin(2C))^2(\bar{\phi}(\pi-2C+\sin(2C))^2-\bar{B}^2\csc^2(\vartheta)\,(2C-2\vartheta-\sin(2C)+\sin(2\vartheta))^2)}\\
&\times[8\cos(\vartheta)-8\cos(2C-\vartheta)-\cos(4C-\vartheta)+16C^2\cos(\vartheta)-32C\vartheta\cos(\vartheta)+16\vartheta^2\cos(\vartheta)-7\cos(3\vartheta)\\
&+\cos(5\vartheta)+8\cos(2C+\vartheta)-\cos(4C+\vartheta)-8Csin(2C-\vartheta)+8\vartheta\sin(2C-\vartheta)+32C\sin(\vartheta)-32\vartheta\sin(\vartheta)\\
&-8Csin(2C+\vartheta)+8\vartheta\sin(2C+\vartheta)]\\
&+\frac{\csc^2(\vartheta)(\pi-2C+\sin(2C))^4\bar{\phi}'^2}{4(\pi-2C+\sin(2C))^2(\bar{\phi}(\pi-2C+\sin(2C))^2-\bar{B}^2\csc^2(\vartheta)\,(2C-2\vartheta-\sin(2C)+\sin(2\vartheta))^2)}
\end{aligned}
\tag{B5}
$$